\documentclass[prd,aps,onecolumn,nopacs,nofootinbib,10pt]{revtex4-2}

\usepackage[english]{babel}
\usepackage[T1]{fontenc}
\usepackage[utf8]{inputenc}
\usepackage{amsmath}
\usepackage{graphicx}
\usepackage{amsfonts}
\usepackage{latexsym}
\usepackage{bbold}
\usepackage{calligra}
\usepackage{ulem}
\usepackage{inputenc}
\usepackage{xspace}
\usepackage{epstopdf}
\usepackage{tikz}
\usepackage{amsthm}
\usepackage{cancel}
\usepackage{comment}
\usepackage[pdftoolbar = true, pdfstartview = FitH, pdfmenubar = true]{hyperref}
\usepackage{stix}
\usepackage{relsize}

\hypersetup{
    colorlinks=true,
    linkcolor=blue,
    filecolor=magenta,      
    citecolor=blue
}

\newcommand{\DI}[1]{{\textbf{DI:} \textcolor{red}{#1}}}

\newcommand{\be}{\begin{equation}}
\newcommand{\ee}{\end{equation}}
\newcommand{\beq}{\begin{equation}}
\newcommand{\eeq}{\end{equation}}
\newcommand{\ba}{\begin{eqnarray}}
\newcommand{\ea}{\end{eqnarray}}

\newcommand{\w}{\mathrm{w}}
\newcommand{\weff}{\mathrm{w}_{\mathrm{eff}}}
\newcommand{\dd}{\mathrm{d}}

\begin{document}

\title{Friedmann cosmology with hyperfluids}

\author{Ilaria Andrei}
\email{ilaria.andrei@ut.ee}
\affiliation{Laboratory of Theoretical Physics, Institute of Physics, University of Tartu, W. Ostwaldi 1, 50411 Tartu, Estonia}

\author{Damianos Iosifidis}
\email{damianos.iosifidis@ut.ee}
\affiliation{Laboratory of Theoretical Physics, Institute of Physics, University of Tartu, W. Ostwaldi 1, 50411 Tartu, Estonia}

\author{Laur Järv} 
\email{laur.jarv@ut.ee}
\affiliation{Laboratory of Theoretical Physics, Institute of Physics, University of Tartu, W. Ostwaldi 1, 50411 Tartu, Estonia}

\author{Margus Saal}
\email{margus.saal@ut.ee}
\affiliation{Laboratory of Theoretical Physics, Institute of Physics, University of Tartu, W. Ostwaldi 1, 50411 Tartu, Estonia}



\begin{abstract} 
In metric-affine gravity, both the gravitational and matter actions depend not just on the metric, but also on the independent affine connection. Thus matter can be modeled as a hyperfluid, characterized by both the energy-momentum and hypermomentum tensors. The latter is defined as the variation of the matter action with respect to the connection and it encodes extra (micro)properties of particles.
For a homogeneous and isotropic universe, it was recently shown that the generic cosmological hypermomentum possesses five degrees of freedom: one in dilation, two in shear, and two in spin part. The aim of the current work is to present the first systematic study of the implications of this perfect hyperfluid on the universe with Friedmann--Lema\^{i}tre--Robertson--Walker metric. We adopt a simple model with non-Riemannian Einstein--Hilbert gravitational action plus arbitrary hyperfluid matter, and solve analytically the cosmological equations for single and multiple component hypermomentum contributions using different assumptions about the equation of state. It is remarkable, that in a number of cases the forms of the time evolution of the Hubble function and energy density still coincide with their general relativity counterparts, only the respective indexes $\weff$ and $\w_\rho$ start to differ due to the hypermomentum corrections. The results and insights we obtained are very general and can assist in constructing interesting models to resolve the issues in standard cosmology.
\end{abstract}

\maketitle
\allowdisplaybreaks
	


\section{Introduction}
\label{intro}

The persistent emergence of puzzles and tensions in contemporary cosmology \cite{Perivolaropoulos:2021jda, Abdalla:2022yfr, DESI:2024mwx}, including the still unknown nature of dark matter and dark energy, motivates us to look beyond the options offered by standard general relativity (GR) \cite{CANTATA:2021asi}. In this regard, a comprehensive framework for a wide range of geometric generalizations of GR is offered by metric-affine gravity (MAG) \cite{Hehl:1976my, Hehl:1976kj, Hehl:1994ue}. While GR uses Riemannian geometry, where the metric completely determines the structure of the manifold and ensures that the connection is automatically torsion-free and metric-compatible, MAG extends this setting to post-Riemanian geometry by treating also the affine (linear) connection as an independent variable.
A generic affine connection can be split into a Levi-Civita connection arising from the metric and an extra tensorial part, 
which is composed of torsion and nonmetricity.
This property allows to formulate MAG as a gauge theory of the four-dimensional affine group, 
where the metric, coframe, and affine connection play the role of gravitational gauge potentials, whereas 
nonmetricity, torsion, and curvature are the corresponding gravitational field strengths
\cite{Lord:1978qz, Gronwald:1997bx, Hehl:1994ue}.
In MAG-type theories, the gravitational Lagrangians are constructed from the metric and the various possible invariants of 
(the combinations of) curvature, torsion, and nonmetricity, while more general Lagrangians of the $f(R)$ kind have been 
considered as well \cite{Sotiriou:2006qn, Vitagliano:2010sr, Capozziello:2007tj, Olmo:2011uz}.
MAG models have been analyzed from the point of view of quantum gravity \cite{Pagani:2015ema}, effective field theory 
\cite{Baldazzi:2021kaf}, cosmological perturbations \cite{Aoki:2023sum}, motion of test particles \cite{Iosifidis:2023eom}, 
conformal invariance \cite{Moon:2009zq, Olmo:2022ops}, and classified according to their scale symmetries 
\cite{Iosifidis:2018zwo, Iosifidis:2019dua}.
In general, higher-order metric-affine models are prone to instabilities \cite{BeltranJimenez:2019acz, BeltranJimenez:2020sqf}, 
but classes free of ghosts and tachyons can still be found \cite{Lin:2018awc, Lin:2020phk, Percacci:2020ddy, Marzo:2021iok, Jimenez-Cano:2022sds, 
Barker:2024ydb, Bahamonde:2024sqo, Bahamonde:2024efl}.


In the MAG framework, the matter fields can also couple to the independent part of the connection, e.g.\ to torsion or nonmetricity. 
This gives rise to a new quantity that has no corresponding notion in the Riemannian theories, namely hypermomentum \cite{Hehl:1976hyperm, Hehl:1976kt, Hehl:1976kv, Hehl:1978cb},
which is formally defined as the variational derivative of the matter part of the action with respect to the affine connection.\footnote{In principle, one can also define the canonical energy-momentum tensor of matter, but there is a certain relation among these three energy-related tensors due to the GL invariance of the matter action
\cite{Hehl:1994ue}.} Therefore in MAG both the usual energy-momentum as well as the hypermomentum of matter act as the sources of gravity \cite{Hehl:1976kj, Hehl:1994ue, Obukhov:2014nja, Puetzfeld:2014qba}, and generate the curvature, torsion, and nonmetricity of spacetime via the metric and connection field equations. 
Using the exterior forms, one can show how the hypermomentum 3-form decomposes into three irreducible pieces of spin (antisymmetric part), dilation (trace part), and proper shear (symmetric traceless part) \cite{Hehl:1994ue}. Here, the idea is similar but should not be confused with the kinematical decomposition of time-like geodesic congruences into vorticity, expansion, and shear parts \cite{Hehl:1971qi}. The most intriguing aspect of the hypermomentum tensor is its association with the micro-properties of matter. For instance, it has been realized  that the antisymmetric part of the affine connection 
(i.e., torsion) is associated with the intrinsic characteristic of matter --- the spin \cite{Hehl:1976kv, Hehl:1976kj}. This means that matter with spin can be a source of torsion in spacetime. It is important to note that these new hypermomentum terms arise naturally from Standard Model fields if we allow them to couple to the connection. Probably the most popular example is that of fermions, which couple to the axial part of torsion and therefore induce a totally antisymmetric hypermomentum \cite{Hehl:1971qi}. In principle, these connection matter couplings can be present for all Standard Model matter fields \cite{Rigouzzo:2023sbb}.


The study of cosmological effects arising in metric-affine gravity has a long history \cite{Puetzfeld:2004rxv}. Most of the work has focused upon the Einstein--Cartan--Sciama--Kibble (EC) theory \cite{Sciama:1964jqa, Kibble:1961ba, Trautman:2004} where the curvature scalar of the Einstein--Hilbert action arises from a generic metric compatible connection, i.e.\ nonmetricity is zero but torsion appears inside matter configurations that carry spin. The first attempt to describe a perfect fluid endowed with intrinsic spin was made by Weyssenhoff and Raabe \cite{Weyssenhoff:1947iua}. Such construction allows a Lagrangian description  \cite{Ray:1982qs, Ray:1982qr, Ray:1982qq, Kopczynski:1986ep, Kopczynski:1990af} and has been considered extensively in the Friedmann--Lema\^itre--Robertson--Walker (FLRW) cosmology for EC gravity \cite{DuartedeOliveira:1986wb, Obukhov:1987yu, Szydlowski:2003nv, Boehmer:2006gd, Brechet:2007cj, Brechet:2008zz, Luz:2018poo, Luz:2023uhy, Medina:2018rnl} including recent attempts to reduce the Hubble tension \cite{Izaurieta:2020xpk, Akhshabi:2023xan}.
The type of spin density with microscopic origin has been also used to generate the inflationary expansion of the early universe within the framework of EC theory \cite{Gasperini:1986mv, Fennelly:1988dx}, although such models require extreme fine-tuning to achieve the required number of e-folds \cite{Gasperini:1986mv}.
It has been also argued that a very small dimensionless density parameter engendering torsion in the context of EC theory offers an alternative solution to the flatness and horizon problems and can thus be a competing model for the inflationary expansion of the early universe \cite{Poplawski:2010kb, Boehmer:2005sw}, or that the inclusion of spin density allows to avoid the initial singularity and restores the usual de Sitter expansion as a particular case \cite{Kopczynski:1972fhu, Demianski:1987hq}.
Recent works embed scalar field inflation into the EC theory, e.g.\ considering the nonminimally coupled Higgs field 
\cite{Rasanen:2018ihz,Shaposhnikov:2020gts,Langvik:2020nrs, Piani:2022gon, He:2023vlj}, or generalizing the Starobinsky model \cite{He:2024wqv}.
The more general metric-affine setups where nonmetricity or higher order terms in the action are also included 
have attracted attention as well for applications in cosmology \cite{Puetzfeld:2001hk, Puetzfeld:2001ur}, 
and more specifically inflation \cite{Akhshabi:2017lyg, Shimada:2018lnm, Mikura:2020qhc, Mikura:2021ldx, 
Rigouzzo:2023sbb,Gialamas:2022xtt, Gialamas:2024jeb, Racioppi:2024zva, racioppi2024tildexi}.

In GR, the study of cosmology generally relies upon the cosmological principle, which postulates that at large scales, the Universe can be considered to be homogeneous and isotropic in space. This restricts the metric and matter energy-momentum tensor to the FLRW form. In the MAG context, it is therefore natural to consider a setup where the post-Riemannian part of the connection, along with the matter hypermomentum tensor, both respect the cosmological principle too. Apart from earlier works of more limited scope \cite{tsamparlis1979cosmological, Minkevich:1998cv, Cabral:2020mzw}, the most general form of spatially homogeneous and isotropic torsion and nonmetricity combined with hypermomentum was presented only recently in Ref.\ \cite{Iosifidis:2020gth}. It was found that the cosmological symmetry allows only five extra geometric variables, two in torsion and three in nonmetricity, while the five hypermomentum degrees of freedom can be represented by two quantities in spin, one in dilaton, and two in shear. The connection equations reduce to the algebraic relations between the connection components and hypermomentum variables, therefore linking torsion to the two spin parts of hypermomentum and the three nonmetricity components to dilation and the two parts of shear. 

The aim of the current paper is to present the first systematic study of the cosmological evolution of FLRW spacetime with hyperfluids. By a cosmological (perfect) hyperfluid we mean matter characterized by the perfect fluid energy-momentum tensor and the aforementioned cosmological form of hypermomentum tensor, while the divergencies of these two tensors are combined into the metric-affine conservation law. Briefly stated, a hyperfluid is an extension of the perfect fluid notion that takes into account the further microscopic characteristics of matter. 
As the cosmological perfect fluid model reduces the microscopic specifics of matter to just the two quantities of density and pressure, the cosmological hyperfluid model does the same with the further microscopic properties, which get reduced to the five quantities of two parts of spin, dilation, and two parts of shear. To solve the cosmological equations in GR we only need to know the density and pressure along with a postulated equation of state that relates those two. Analogously, in MAG cosmology, we need to know in addition only the spin, dilation, and shear quantities, and can integrate the field equations after postulating some equation of state for these hypermomentum variables (e.g.\ that spin etc., are proportional to the density).
Thus, the approach we take is completely general, and does not require assuming any particular form of the microscopic matter Lagrangian. Given any Lagrangian where the matter fields couple to the post-Riemannian connection, hypermomentum 
can be immediately derived, and the results of our analysis apply.
Just to be clear, the Weyssenhoff fluid with spin is incompatible with the cosmological principle \cite{Boehmer:2006gd} and does not fall under the purview of the present work. 
Also, the previous hyperfluid models  \cite{Obukhov:1993pt, Obukhov:1996mg, Babourova:1998mgh, Obukhov:2023yti} are different as well, since in these constructions one can not generate all the hypermomentum pieces that are allowed by the cosmological principle (see \cite{Boehmer:2006gd} for a relevant discussion). Some aspects of the cosmological hyperfluid we are studying here have been considered in the recent works \cite{Iosifidis:2020upr, Iosifidis:2021fnq, Iosifidis:2022evi, Iosifidis:2024bsq}.


We consider the most straightforward MAG generalization of GR, where the gravitational action is given by the curvature scalar that depends on the generic affine connection, while the matter action involves couplings to the generic affine connection too. Since the curvature scalar is invariant under projective transformations of the connection, the dilaton current in the matter sector must vanish. Furthermore, in our model, as we explicitly show, one of the cosmological spin modes can be immediately integrated to behave as spatial curvature, and we will drop it from further analysis. Hence the cosmological hyperfluid in this theory reduces to one spin and two shear components. The EC gravity \cite{Sciama:1964jqa, Kibble:1961ba, Trautman:2004} is covered as a particular subclass of our model corresponding to vanishing shear. The model reduces to GR in vacuum or when the matter carries no hypermomentum.

The paper is organized as follows. First, in Section \ref{sec: 2 Metric-affine extension of general relativity}, we describe the general formalism and all the conventions used. We introduce the field equations in the presence of both torsion and nonmetricity, and establish the relevant conservation laws. Then, in Sec.\ \ref{sec: 3 Spatially flat cosmology}, we focus upon the consequences of cosmological principle on the connection and hypermomentum, and discuss how to approach the final set of cosmological equations. After briefly recalling, in Sec.\ \ref{GR}, how the cosmological solutions arise in GR, we turn to the single component hyperfluids in Sec.\ \ref{sec: 5 Single component hypermomentum} and consider the three cases of spin and two parts of shear in detail. We show that under various assumptions about the hypermomentum equation of state, the cosmological solutions turn out to be similar to the ones in GR, only the barotropic index that describes the dilution of the energy density and the effective index that describes the expansion of the Universe get modified. We tackle some simple examples of multicomponent hyperfluid cosmologies in Sec.\ \ref{sec: 6 Multiple component hyperfluid}, 
and conclude with a summary and deliberation on the prospects for model building in Sec.\ \ref{sec: 7 Conclusions}. In the end, the Appendix \ref{sec: appendix} presents the relations between the hypermomentum variables.

\section{Metric-affine extension of general relativity}
\label{sec: 2 Metric-affine extension of general relativity}

\subsection{Geometric preliminaries}

Let us briefly introduce here some basic definitions and conventions we shall be using throughout. 
We adopt the ones that appear in \cite{Iosifidis:2019dua}.  
We consider a general 4-dimensional non-Riemannian space, endowed with a metric $g_{\mu\nu}$ and generic affine 
connection with coefficients $\Gamma^{\lambda}{}_{\mu\nu}$ decomposed as
\begin{equation}
\label{decgamma}
    \Gamma^{\lambda}{}_{\mu \nu} = \tilde{\Gamma}^{\lambda}{}_{\mu \nu} + N^{\lambda}{}_{\mu \nu}\,,
\end{equation}
where
\beq
\label{lcconn}
    \tilde{\Gamma}^\lambda_{\phantom{\lambda}\mu\nu} = \frac{1}{2} g^{\rho\lambda}
    \left(\partial_\mu g_{\nu\rho} + \partial_\nu g_{\rho\mu} - \partial_\rho g_{\mu\nu}\right)
\eeq
is the usual Levi-Civita connection, and the tensor
\beq
\label{distortion}
    {N^\lambda}_{\mu\nu} = {\frac12 g^{\rho\lambda}\left(Q_{\mu\nu\rho} + Q_{\nu\rho\mu}
    - Q_{\rho\mu\nu}\right)} - {g^{\rho\lambda}\left(S_{\rho\mu\nu} +
    S_{\rho\nu\mu} - S_{\mu\nu\rho}\right)} \,
\eeq	
is called distortion, as it quantifies the departure from the Riemannian geometry. 
The symmetric sector of the post-Riemannian part of the affine connection is characterized by nonmetricity 
\beq
    Q_{\alpha\mu\nu}:=-\nabla_{\alpha}g_{\mu\nu}
    = \partial_{\alpha}g_{\mu\nu} + \Gamma^{\lambda}_{\phantom{\lambda}\mu\alpha}g_{\lambda\nu}
    +\Gamma^{\lambda}_{\phantom{\lambda} \nu\alpha}g_{\lambda\mu}
\eeq
and the antisymmetric sector by torsion
\beq
    S_{\mu\nu}^{\phantom{\mu\nu}\lambda}:=\Gamma^{\lambda}_{\phantom{\lambda}[\mu\nu]} \,.
\eeq
Even though it may sound peculiar, the symmetric part of the connection (in the last two indices) has torsion as well. 
Namely $N^{\lambda}{}_{(\mu\nu)}$ has torsion. On the other hand, $N_{(\alpha\mu)\nu}$ is free from torsion.

Taking contractions from nonmetricity, we can obtain two vectors
\beq
    Q_{\mu}:=Q_{\mu\alpha \beta}g^{\alpha \beta}\,, \qquad 
    \bar{Q}_{\mu}:=Q_{\alpha\nu\mu}g^{\alpha\nu}
\eeq
with the one at the left oftentimes referred to as the Weyl vector. 
As for torsion, we can define a vector along with a pseudo-vector as follows
\beq
    S_{\mu}:=S_{\mu\alpha}^{\phantom{\mu\alpha} \alpha}\,, \qquad t_{\mu}:=\epsilon_{\mu\nu\alpha \beta}S^{\nu\alpha \beta} \,,
\eeq
the latter being a parity-odd piece. 
In addition, the connection is also characterized by the curvature
\begin{equation}
    R^{\mu}_{\phantom{\mu} \nu \alpha \beta}:= 
    2\partial_{[\alpha}\Gamma^{\mu}_{\phantom{\mu}|\nu|b]}+2\Gamma^{\mu}_{\phantom{\mu}\rho[\alpha}\Gamma^{\rho}_{\phantom{\rho}|\nu|b]} \,.
\end{equation}
Here, horizontal bars around an index denote that this index is left out of the (anti)-symmetrization. 
In a general metric-affine setting, the curvature tensor is antisymmetric in its last two indices as follows immediately 
by the above definition.  
Without the use of metric, we can construct the two independent contractions for the curvature tensor
\beq \label{Ricci tensor}
    R_{\nu \beta}:=R^{\mu}_{\phantom{\mu} \nu \mu \beta}\,, \qquad 
    \widehat{R}_{\alpha \beta}:=R^{\mu}_{\phantom{\mu} \mu \alpha \beta} \,.
\eeq
The former defines, as usual, the Ricci tensor, while the latter is the so-called homothetic curvature tensor 
and is of purely post-Riemannian origin\footnote{In particular, it can be easily shown to be the field 
strength of $\Gamma^{\lambda}_{\phantom{\lambda}\lambda\mu}$.}. 
Once a metric is given, we can form yet another  contraction
\beq \label{co-Ricci tensor}
    \breve{R}^{\lambda}_{\phantom{\lambda} \kappa}:=R^{\lambda}_{\phantom{\lambda} \mu\nu\kappa}g^{\mu\nu} \,,
\eeq
which is oftentimes referred to as the co-Ricci tensor. However, the Ricci scalar is still uniquely defined since
\beq \label{Ricci scalar}
    R:=R_{\mu\nu}g^{\mu\nu}=-\breve{R}_{\mu\nu}g^{\mu\nu}\,, \qquad  \widehat{R}_{\mu\nu}g^{\mu\nu}=0 \,.
\eeq
The above are the main geometric ingredients needed for the rest of our analysis. We may now focus on the physical aspects.

\subsection{The Matter sector of MAG}

In MAG, apart from the usual metrical energy-momentum tensor, defined as usual by
\begin{align}
    T^{\alpha \beta} :=+\frac{2}{\sqrt{-g}}\frac{\delta(\sqrt{-g} \mathcal{L}_{M})}{\delta g_{\alpha \beta}} \,,
\end{align}
we also have a new energy-related object called the hypermomentum \cite{Hehl:1976hyperm, Hehl:1978cb},
which is defined as the variation of the matter part of the action with respect to the independent affine connection, viz.
\begin{align}
    \Delta_{\lambda}^{\phantom{\lambda} \mu\nu} 
   : = -\frac{2}{\sqrt{-g}}\frac{\delta ( \sqrt{-g} \mathcal{L}_{M})}{\delta \Gamma^{\lambda}_{\phantom{\lambda}\mu\nu}} \,.
\end{align}
These tensors are not totally independent, but they are related by 
\cite{Iosifidis:2020gth, Obukhov:2013ona, Obukhov:2014nja}
\beq
    \sqrt{-g}(2 \tilde{\nabla}_{\mu}T^{\mu}_{\;\;\alpha}-
    \Delta^{\lambda\mu\nu}R_{\lambda\mu\nu\alpha})+\hat{\nabla}_{\mu}\hat{\nabla}_{\nu}(\sqrt{g}
    \Delta_{\alpha}^{\;\;\mu\nu})+2 S_{\mu\alpha}^{\;\;\;\;\lambda} \, \hat{\nabla}_{\nu}(\sqrt{-g}
    \Delta_{\lambda}^{\;\;\;\mu\nu})=0 \,,
\label{ccc}
\eeq
which comes from the diffeomorphism invariance of the matter sector.
In the above $\tilde{\nabla}$ is the Levi-Civita covariant derivative and
\beq
    \hat{\nabla}_{\nu}:= 2 S_{\nu}-\nabla_{\nu} \,.
\eeq
The expression \eqref{ccc} serves as the generalization of the conservation law of energy-momentum for matter with 
microstructure. In the differential forms formulation, the hypermomentum form is split into its three irreducible 
pieces of spin, dilation, and shear. 
Translated in the coordinate formalism, the hypermomentum decomposition reads
\beq
    \Delta_{\mu\nu\alpha}=\tau_{\mu\nu\alpha}+\frac{1}{4}\Delta_{\alpha} g_{\mu\nu}+\hat{\Delta}_{\mu\nu\alpha}
\eeq
with the spin (antisymmetric), dilation (trace), and shear (symmetric traceless) quantities, respectively, given by
\begin{subequations}
\begin{align}
    \tau_{\mu\nu\alpha} &= \Delta_{[\mu\nu]\alpha} \,, \\
    \Delta_{\alpha} &= \Delta_{\mu\nu\alpha}g^{\mu\nu} \,, \\
    \hat{\Delta}_{\mu\nu\alpha} &= \Delta_{(\mu\nu)\alpha}-\frac{1}{4}\Delta_{\alpha} g_{\mu\nu} \,.
\end{align}
\end{subequations} 
Having also defined the relevant physical quantities, let us proceed with the construction of the model we shall 
analyze in detail.

\subsection{Metric-affine gravity model}

 Our action will consist of the usual Einstein--Hilbert term along with a matter sector of a perfect hyperfluid 
 encoding the microstructure of matter. That is, our theory is given by
\begin{align}
\label{S}
    S=\frac{1}{2 \kappa}\int \mathrm{d}^{4}x \sqrt{-g} \left[ R(g_{\mu\nu}, \, \Gamma^\lambda{}_{\mu\nu}) 
    + \mathcal{L}_M \, (g_{\mu\nu}, \, \Gamma^\lambda{}_{\mu\nu}, \, \chi_M) \right] \,,
\end{align}
where $R$ is the Ricci scalar \eqref{Ricci scalar} computed from the general connection \eqref{decgamma}, 
and $\mathcal{L}_M$ is the Lagrangian of the matter fields $\chi_M$.
It is well-known and quite trivial to show that the generalized scalar curvature $R$ is projective invariant. 
That is, the Einstein--Hilbert term is invariant under connection transformations of the form
\beq
\label{projective}
    \Gamma^{\lambda}{}_{\mu\nu} \mapsto \Gamma^{\lambda}{}_{\mu\nu}+\delta^{\lambda}_{\mu}\xi_{\nu} \,,
\eeq
while keeping the metric tensor $g_{\mu\nu}$ intact. 
In the above,  $\xi_{\nu}$ is an arbitrary 1-form. 
This implies that the so-called Palatini tensor, defined as 
\beq
    P_{\lambda}{}^{\mu\nu}:=\frac{\delta R}{\delta\Gamma^{\lambda}{}_{\mu\nu} } \,,
\eeq
is traceless in its first tow indices, namely $P_{\lambda}{}^{\lambda\nu} \equiv 0$.
This fact, by means of the connection field equations, implies then that the dilation current must be vanishing:
\beq
    \Delta_{\lambda}{}^{\lambda\nu}=0 \,.
\eeq
As a result, the choice of the gravitational action constraints the form of the hypermomentum to be that of 
a dilation-free type, which is what we shall be dealing with from now on\footnote{The most natural way to alleviate 
this constraint (i.e.\ have a non-vanishing dilation) is to supplement the Einstein--Hilbert term with quadratic 
invariants in torsion and/or nonmetricity. }. 

Varying the action \eqref{S} with respect to the metric and the connection gives us the Einstein and Palatini field 
equations
\begin{align}
    R_{(\mu\nu)}-\frac{1}{2}g_{\mu\nu}R &= \kappa T_{\mu\nu} \,, \label{metrf} \\
    \left( \frac{Q_{\lambda}}{2}+2 S_{\lambda}\right) g^{\mu\nu}-(Q_{\lambda}{}^{\mu\nu}
    +2 S_{\lambda}{}^{\mu\nu})+\left( \bar{Q}^{\mu}-\frac{Q^{\mu}}{2}-2 S^{\mu} \right)\delta^{\nu}_{\lambda} 
    &=\kappa \Delta_{\lambda}^{\phantom{\lambda} \mu\nu} \,. \label{conf}
\end{align}

In the case when the matter Lagrangian does not depend on the post-Riemannian part of the connection and hypermomentum vanishes, 
the conservation law \eqref{ccc} reduces to the usual $\tilde{\nabla}_{\mu}T^{\mu}_{\;\;\alpha}=0$, while the connection equation
restricts the torsion and nonmetricity to pure gauge i.e.\ they can be turned to zero by projective freedom \eqref{projective}, 
and what remains is the Levi-Civita connection and Einstein's equation of general relativity. Similarly, if we start off with 
no shear, the nonmetricity can be made to vanish, only the torsional components of the post-Riemannian connection remain, 
and the theory reduces to Einstein--Cartan--Sciama--Kibble gravity \cite{Sciama:1964jqa, Kibble:1961ba, Trautman:2004}. 
Zero shear is also trivially consistent with Weyl--Cartan geometry which in general, has torsion and only the trace part of 
nonmetricity, but the latter vanishes here due to projective invariance.
Finally, if spin is taken to be identically zero, there will be no torsion, giving a theory that might be called Palatini.

\section{Spatially flat cosmology}
\label{sec: 3 Spatially flat cosmology}

\subsection{Cosmological geometry}

Imposing that the metric is homogeneous and isotropic and its spatial curvature vanishes gives the well-known FLRW metric 
after choosing the coordinates suitably,
\begin{align}
    \mathrm{d}s^{2}=-\mathrm{d}t^{2} + a^{2}(t)\delta_{ij} \mathrm{d}x^{i} \mathrm{d}x^{j} \,.
\end{align}
Here $a(t)$ is a single free function that measures the expansion of space, called the scale factor. If we impose the same 
symmetry on a general affine connection \eqref{decgamma}, the Levi-Civita part obeys it by default, while  the distortion 
tensor reads in the covariant form  as \cite{Iosifidis:2020gth}
\begin{gather} 
\label{distortion FLRW}
N_{\alpha\mu\nu} = X(t)\, u_{\alpha}h_{\mu\nu} + Y(t)\, u_{\mu}h_{\alpha\nu} + Z(t) \, u_{\nu}h_{\alpha\mu} + 
V(t)\, u_{\alpha}u_{\mu}u_{\nu} + \epsilon_{\alpha\mu\nu\lambda}\, u^{\lambda} W(t) \,,
\end{gather}
where $X(t)$, $Y(t)$, $Z(t)$, $V(t)$, and $W(t)$ are five additional functional degrees of freedom compatible with 
the cosmological symmetry. 
In the above $u^{\mu}$ is the fluid's 4-velocity field, normalized as $u_\mu u^\mu=-1$ and which, in co-moving coordinates, 
is expressed as $u^\mu= \delta^\mu_0=(1,0,0,0)$. Moreover, 
\beq
\label{projop}
    h_{\mu \nu}:= g_{\mu \nu} + u_\mu u_\nu \,,
\eeq
is the usual projection tensor.

Alternatively, one can impose the symmetry on the torsion and nonmetricity tensors directly, to discover that torsion 
contributes two degrees \cite{tsamparlis1979cosmological} of freedom (in $4$-dimensions) while nonmetricity contributes 
three degrees of freedom \cite{Minkevich:1998cv}. 
The covariant form of the torsion and nonmetricity tensors that respect spatial isotropy and homogeneity, 
then read \cite{Iosifidis:2020gth}
\beq
    S_{\mu\nu\alpha}=2 u_{[\mu}h_{\nu]\alpha}\Phi(t)+\epsilon_{\mu\nu\alpha\rho}u^{\rho} P(t) 
\label{isotor}
\eeq
and 
\beq 
\label{isononmet}
    Q_{\alpha \mu \nu}  = A(t) u_\alpha h_{\mu \nu} + B(t) h_{\alpha(\mu} u_{\nu)} + C(t) u_\alpha u_\mu u_\nu \,,
\eeq
respectively. Here $\Phi, P$ are the degrees of freedom of torsion while $A, B$, and $C$ are the three degrees of freedom 
fully describing the evolution of nonmetricity in such a setting. 
Then, using the relations
\beq
    Q_{\nu\alpha\mu}=2 N_{(\alpha\mu)\nu} \ , \qquad
    S_{\mu\nu\alpha}=N_{\alpha[\mu\nu]} \,,
\eeq	
one can establish a link between the torsion and nonmetricity variables and the distortion degrees of freedom according to 
\beq
    2(X+Y) = B \ , \qquad  2 Z = A \ , \qquad  2 V = C \ , \qquad  2 \Phi = Y-Z \ , \qquad  P = W \ ,	 
\label{dv}
\eeq
with inverse form
\beq
    W = P \ , \qquad  V = \frac{C}{2} \ , \qquad Z=\frac{A}{2} \ , \qquad Y=2\Phi +\frac{A}{2} \ , 
    \qquad X=\frac{B}{2} - 2 \Phi -\frac{A}{2} \ .
\label{V C}
\eeq
Whichever set one uses is then totally irrelevant to the actual physics; the first set (i.e. torsion/nonmetricity 
variables) 
has a more transparent geometrical interpretation, while the latter (i.e.\ distortion variables) is oftentimes more 
convenient for actual calculations. 

\subsection{Cosmological matter}

To obtain consistent solutions, we must impose the same symmetry on the matter sector as well. 
The energy-momentum tensor of matter in an FLRW spacetime necessarily has the usual perfect fluid form
\beq
    T_{\mu\nu} = \rho \, u_{\mu}u_{\nu} + p \, h_{\mu\nu} \,,
\label{metrical} 
\eeq
where $\rho$ and $p$ are the usual density and pressure of the matter and
$h_{\mu\nu}$ the projection tensor \eqref{projop}. 
Similarly, under the cosmological symmetry, the hypermomentum tensor takes the covariant form \cite{Iosifidis:2020gth}	
\beq 
\label{hypermomentum:degrees}
    \Delta_{\alpha\mu\nu} = \phi \, h_{\mu\alpha}u_{\nu} + \chi \, h_{\nu\alpha}u_{\mu} + \psi \, u_{\alpha}h_{\mu\nu} 
    + \omega \, u_{\alpha}u_{\mu} u_{\nu} + \epsilon_{\alpha\mu\nu\kappa}\, u^{\kappa}\zeta \,,
\eeq
where $\phi, \chi, \psi, \omega, \zeta$ are the $5$ matter variables describing the hypermomentum part of the cosmological fluid.

To give a more physical interpretation, the remaining five fields can be rearranged to match the spin, dilation, 
and shear parts of hypermomentum according to (see \cite{Hehl:1994ue}) 
\begin{subequations}
\begin{align}
    \tau_{\alpha\mu\nu} & = (\psi-\chi) \, u_{[\alpha}h_{\mu]\nu} + \epsilon_{\alpha\mu\nu\kappa}u^{\kappa}\zeta  \,, \\
    \Delta_{\nu} &= (3 \phi -\omega) \, u_{\nu} \,, \\
    \hat{\Delta}_{\alpha\mu\nu} &= \frac{(\phi+\omega)}{4}\left[ h_{\alpha\mu}+3 u_{\alpha}u_{\mu} \right] u_{\nu} 
    + (\psi +\chi) \, u_{(\mu}h_{\alpha)\nu} \,.
\end{align}
\end{subequations}
As we can see, the fields $\psi, \chi, \phi, \omega$ 
have no direct physical interpretation just by themselves, 
but rather specific combinations of them correspond to the spin, dilation, and shear of matter. 
It is helpful to define 
\begin{subequations}
\label{sigma psi}
\begin{align}
    \sigma &= \frac{(\psi-\chi)}{2}  \hspace{1 cm}  \zeta =\zeta
    \label{spinsig} \\ 
    \Delta &= 3 \phi-\omega \,,  \label{dilphiomega}\\	
    \Sigma_{1} &= \frac{(\psi+\chi)}{2} \,, \qquad \Sigma_{2}=\frac{(\phi+\omega)}{4}
    \label{shearSig} 
\end{align}
\end{subequations}
where the first line contains spin degrees of freedom, the second relates to the dilation, and the third correspond to the two parts of shear $\Sigma_{1}, \Sigma_{2}$, see Appendix \ref{sec: appendix}.
With these redefinitions, the three sources of the hypermomentum take the form
\begin{subequations}
\begin{align}
    \tau_{\alpha\mu\nu} &= 2 \sigma \, u_{[\alpha}h_{\mu]\nu} + \epsilon_{\alpha\mu\nu\kappa}\, u^{\kappa}\zeta \,, \\ 
    \Delta_{\nu} &= \Delta \, u_{\nu}   
\label{dil} \,, \\
    \hat{\Delta}_{\alpha\mu\nu} &= \Sigma_{2} \, \left[ h_{\alpha\mu} + 3 u_{\alpha}u_{\mu} \right] u_{\nu} 
    + 2 \Sigma_{1} \, u_{(\mu}h_{\alpha)\nu} \,.
\end{align}
\end{subequations} 
The inverse relations read
\beq
    \psi = \Sigma_{1} + \sigma \ ,  \qquad	\chi = \Sigma_{1} - \sigma \ , \qquad  
    \phi = \Sigma_{2} + \frac{1}{4}\Delta \ , \qquad  \omega=3\Sigma_{2}-\frac{1}{4}\Delta \,, \qquad  \zeta=\zeta \,. 
\eeq

\subsection{Torsion and nonmetricity in terms of their sources}

By invoking the connection (Palatini) equation \eqref{conf}, we can express the torsion and nonmetricity variables 
in terms of their sources (hypermomentum pieces). 
Indeed, by substituting the cosmological torsion \eqref{isotor} and nonmetricity \eqref{isononmet} into \eqref{conf}, 
we find
\begin{align}
    \left[\frac{A}{2} +4\Phi -\frac{C}{2}\right] u_{\alpha}h_{\mu\nu}
    +\left[B-\frac{3A}{2} -4\Phi -\frac{C}{2}\right] u_{\mu}h_{\alpha\nu} 
    -\frac{B}{2}u_{\nu}h_{\mu\alpha}-\frac{3B}{2}u_{\alpha}u_{\mu}u_{\nu}
    -2\epsilon_{\alpha\mu\nu\rho}u^{\rho}P 
    &= \kappa \Delta_{\alpha \mu\nu} \,,
\end{align}
where the hypermomentum $\Delta_{\alpha \mu\nu}$ is given by \eqref{hypermomentum:degrees}. 
Equating pieces with similar structures gives relations
\begin{subequations}
\label{A psi}
\begin{align}
\label{A:psi}
    \frac{A}{2}+4\Phi -\frac{C}{2} &= \kappa \psi \,, \\
\label{A:chi}
    B-\frac{3A}{2} -4\Phi -\frac{C}{2} &= \kappa \chi \,, \\
    2 P &= -\kappa \zeta \,, \\
\label{B:phi}    
    B &= -2 \kappa \phi \,, \\
\label{B:omega}    
    3B &= - 2 \kappa \omega \,.
\end{align}
\end{subequations}
Notice that the latter two relations imply the projective invariance constraint on the matter variables, namely that the dilaton 
\eqref{dilphiomega}, \eqref{dil} vanishes and hence
\begin{align}
\label{condition:dilation:zero}
    \omega = 3 \phi \,.
\end{align}

Additionally, we can fix the gauge (projective freedom) in such a way so as to obtain a vanishing 
$\bar{Q}_{\mu}$,  then $\bar{Q}_{\mu}=0$ implies 
\beq
    C = \frac{3 B}{2} \,.
\eeq	
We see that the sources of torsion are $\psi$ and $\chi$, along with $\zeta$, while for nonmetricity, we have 
$\phi$, $\chi$, $\psi$, $\omega$. 
The above derivation was general, and no additional constraints on the hypermomentum were imposed (apart from the 
off-shell vanishing dilation). 

\subsection{Matter continuity equation}

In terms of the quantities 
\beq 
\label{F}
    F = \dot{\Sigma}_{2} +3 H \Sigma_{2}+ 2\Sigma_{1}(H-Y)+\frac{B}{2}(\Sigma_{1}+\sigma) 
\eeq
and
\beq
    D = \dot{\Delta}+3 H \Delta \,,
\eeq
the continuity equation \eqref{ccc} appears as
\begin{align}
    \Big[\dot{\rho}+3H(\rho +p)\Big]+\frac{3}{2}(\dot{F}+4HF)-\frac{1}{8}\dot{D} &= \, \sigma R_{\mu\nu}\, u^{\mu}u^{\nu} 
    +(\Sigma_{1}+\sigma)( R_{\mu\nu}+ \breve{R}_{\mu\nu}) \, u^{\mu}u^{\nu} \nonumber \\
    & +\frac{3}{4}(A+C)F + \frac{1}{16} ( 3A - C )D \,,
\end{align}
where the Ricci \eqref{Ricci tensor} and co-Ricci \eqref{co-Ricci tensor} tensors for the cosmological 
connection \eqref{distortion FLRW} are (see \cite{Iosifidis:2020upr}) 
\begin{gather}
    R_{\mu\nu}= \Big[\dot{X}+H \big( 3X-2Y+(Z+V) \big) -2X Y+X(Z+V)-2 W^{2}+\dot{H}+3H^{2}\Big] \,  h_{\mu\nu}
    \nonumber \\
    +\Big[ \dot{Y}+H(Y+Z+V)-Y(Z+V)-(\dot{H}+H^{2})\Big]\, 3u_{\mu}u_{\nu}
\end{gather}
and 
\begin{gather}
    \breve{R}_{\mu\nu}=\Big[ \dot{Y}+H \big(-2X+3Y+(Z+V) \big)+2X Y -Y (Z+V) + 2 W^{2} -\dot{H}-3\dot{H}^{2}\Big]h_{\mu\nu} 
    \nonumber \\
    +\Big[ \dot{X}+H (X+Z+V)+X (Z+V)+(\dot{H}+H^{2}) \Big] \, 3u_{\mu}u_{\nu} \,.
\end{gather}
It makes sense to express the continuity equation in a single set of variables. Thus using the relations 
\eqref{V C}, \eqref{sigma psi}, \eqref{A psi} we can write all in terms of the spin (\ref{spinsig}) and 
shear (\ref{shearSig}) components of hypermomentum
\beq
    \Big[ \dot{\rho} + 3 H(\rho+p)\Big]+\frac{3}{2}\Big[ \dot{F} + 4 H F \Big]
    =\frac{\kappa}{2}(\rho + 3 p)\sigma +\frac{3 \kappa}{2}F(\sigma-\Sigma_{2}) \,.
\label{contnocan}
\eeq
From \eqref{B:phi} we get
\begin{equation}
\label{B:Sigma2}
    B = - 2 \kappa \Sigma_{2} \,
\end{equation}
and with this we find
\begin{equation}
\label{Y:sigma:Sigma2}
    Y = \frac{1}{2} \kappa (\sigma  - \Sigma_{2}) \,
\end{equation}
and subsequently the function $F$ takes the on-shell form
\begin{equation}
\label{F:sigmas}
    F = \dot{\Sigma}_{2} + 3 H (\Sigma_2 + \frac{2}{3} \Sigma_1) - \kappa \sigma (\Sigma_1 + \Sigma_2) \,.
\end{equation}
Note that Eq.\ \eqref{contnocan} is the most general continuity equation that relates the energy density of ordinary matter $\rho$, 
the spin $\sigma$, and shear $\Sigma_{1}$,  $\Sigma_2$ (recall that dilation vanishes in our case) and takes into account the 
cosmological principle.

\subsection{Metric field equations}

From the metric field equations, we extract 
\begin{gather}
    H^{2}=-H\left[ \frac{3}{2}X-\frac{1}{2}Y+Z+V \right] -\frac{1}{2}(\dot{X}+\dot{Y})-\frac{1}{2}(X-Y)(Z+V)+XY 
    +\frac{\kappa}{3}\rho \,,
\end{gather}
which is the first modified Friedmann equation. 
Then, employing the connection field equations, we may eliminate the distortion variables in favor of the 
hypermomentum sources to arrive at
\beq
    H^{2}=\frac{\kappa}{2}\left( \dot{\phi}+(3\phi + \psi+\chi)H-\frac{\kappa}{4}(\psi-\chi)(\psi+\chi+2 \phi)\right) 
    +\frac{\kappa^{2}}{4}\phi^{2}+\frac{\kappa^{2}}{4}\zeta^{2}-\frac{\kappa^{2}}{16}(\psi-\chi)^{2}
    +\frac{\kappa}{2}H(\psi-\chi)+\frac{\kappa \rho}{3} \,.
\eeq
In addition, from the metric field equations, we can also obtain the modified acceleration equation which, after expressing 
everything in terms of the hyperfluid variables, reads
\begin{gather}
    \frac{\ddot{a}}{a}=-\frac{\kappa}{6}(\rho + 3 p)+\frac{\kappa}{4}(\dot{\psi}-\dot{\chi})+\frac{\kappa}{4}H(\psi - \chi)
    -\frac{\kappa^{2}}{4}\phi (\psi +\chi+ 2 \phi) 
    -\frac{\kappa}{2}\left[ \dot{\phi}+ (3\phi +\chi +\psi)H -\frac{\kappa}{4}(\psi-\chi)(\psi+\chi+2 \phi)\right] \,.
\end{gather}
Finally, in terms of the spin and shear quantities, the metric equations read \cite{Iosifidis:2020gth}
\begin{subequations}
\begin{align}    
\label{Fr1 eq}
    H^{2} &=\frac{\kappa}{3}\rho +\frac{\kappa}{2}F+\frac{\kappa^{2}}{4}\Sigma_{2}^{2}-\frac{\kappa^{2}}{4}\sigma^{2}
    +\kappa \sigma H +\frac{\kappa^{2}}{4}\zeta^{2}\,, \\
\label{Fr2 eq}
    \frac{\ddot{a}}{a} &=-\frac{\kappa}{6}(\rho + 3 p)-\frac{\kappa}{2}F+\frac{\kappa}{2}(\dot{\sigma}+H\sigma)
    -\frac{\kappa^{2}}{2}\Sigma_{2}(\Sigma_{1}+\Sigma_{2}) \,,
\end{align} 
\end{subequations} 
where $F$ has the form given in \eqref{F:sigmas}.

One can check that out of the three equations \eqref{Fr1 eq}, \eqref{Fr2 eq}, \eqref{contnocan}, only two are 
independent provided (see also \cite{Iosifidis:2024bsq})
\begin{align}
\label{eq zeta eq}
    \left(\dot{\zeta} + H \zeta \right) \zeta &= 0.
\end{align}
Thus the consistency of the equations yields an extra condition for $\zeta$ that immediately fixes $\zeta = 0$ or $\zeta \sim a^{-1}$. 
The latter means the $\zeta^2$ term in the Friedmann equation \eqref{Fr1 eq} behaves exactly as the spatial curvature $k$ term would behave
and completely decouples from the rest of the hyperfluid variables. Therefore, since the effect of a curvature term in the Friedmann 
equations is well known, from now on, we shall set $\zeta=0$ and focus on the rest of the hyperfluid variables. 
We must stress however that this result depends crucially upon our choice for the gravitational action and for more general actions 
(i.e.\ including also quadratic torsion and nonmetricity invariants) the situation will change drastically. 
Finally, it is worth mentioning that, had we started with a non-vanishing curvature term (of either sign) the presence of $\zeta$ 
could screen this curvature and effectively give a vanishing total curvature. This is an interesting feature that might be worth exploring further. 

\subsection{System of cosmological equations}

We can write the field equations \eqref{Fr1 eq} (with $\zeta=0$), \eqref{Fr2 eq}, \eqref{contnocan} as
\begin{subequations}
\label{eq: FLRW equations general}
\begin{align}
\label{eq: FR1}
    3 H^2 &= \kappa \rho + \kappa \rho_h \,, \\
\label{eq: FR2}
    2 \dot{H} + 3 H^2 &= - \kappa p - \kappa p_h \,, \\
\label{eq: continuity eq}
    \dot{\rho} + 3 H (\rho + p ) &= -\dot{\rho}_h - 3 H (\rho_h + p_h ) \,,
\end{align}
\end{subequations}
where the effective density and pressure of the hyperfluid is given by
\begin{align}\label{rhohph}
    \rho_h &:= \frac{3 \dot{\Sigma}_2}{2} + \kappa \left(- \frac{3 \Sigma_{1} \sigma}{2} + \frac{3 \Sigma_{2}^{2}}{4} 
    - \frac{3 \Sigma_{2} \sigma}{2} - \frac{3 \sigma^{2}}{4}\right) + H \left(3 \Sigma_{1} + \frac{9 \Sigma_{2}}{2} 
    + 3 \sigma\right)\,, \\
    p_h &:= \frac{\dot{\Sigma}_2}{2} - \dot{\sigma} + \kappa \left(\Sigma_{1} \Sigma_{2} - \frac{\Sigma_{1} \sigma}{2} 
    + \frac{3 \Sigma_{2}^{2}}{4} - \frac{\Sigma_{2} \sigma}{2} + \frac{\sigma^{2}}{4}\right) + H \left(\Sigma_{1} 
    + \frac{3 \Sigma_{2}}{2} - 2 \sigma\right) \,. \label{pph}
\end{align}
Note that only two of the equations \eqref{eq: FLRW equations general} are independent since either of the last two 
can be derived from the remaining two and the time derivative of the first. 
We can also define barotropic indexes of the components as well as effectively for the total system as
\begin{subequations}
\begin{align}
\label{eq: w definition}
    \w &:=\frac{p}{\rho} \,, \\
\label{eq: w_h definition}
    \w_h &:= \frac{p_h}{\rho_h} \,, \\
\label{eq: w_eff definition general}
    \w_{\mathrm{eff}} & := \frac{p + p_h}{\rho + \rho_h} = \frac{\w \, \rho + \w_h \, \rho_h}{\rho + \rho_h} \neq \w + \w_h \,.
\end{align}
\end{subequations}
With our equations, it follows that
\begin{equation}\label{eq: w_eff definition}
     \w_{\mathrm{eff}} =  - 1 - \frac{2 \dot{H}}{3H^2} \,.
\end{equation}
We name these quantities in the following way: $\w$ as a barotropic index of matter, $\w_h$ as a hypermomentum index, 
and $\w_{\mathrm{eff}}$ as effective index.
In the system \eqref{eq: FLRW equations general} we have six unknown functions $\rho(t)$, $p(t)$, $\sigma(t)$, $\Sigma_1(t)$, $\Sigma_2(t)$, $H(t)$ and only two independent equations. In order to solve the system, we have to make some assumptions which impose four constraints among these variables. Such under-determination is not something extraordinary, as in essence, it appears already in the GR cosmology where the hyperfluid components vanish, and solving the system needs fixing the relationship between the density $\rho(t)$ and pressure $p(t)$ of matter in terms of the barotropic index $\w$. The barotropic index can be derived from the microscopic properties of the matter fluid one wants to consider (radiation, dust, etc.), i.e.\ from the fundamental Lagrangian of the matter fluid. In principle, many different fundamental Lagrangians can give the same barotropic index, and therefore, in the study of cosmological evolution, it is not necessary to fix the matter Lagrangian precisely. Just setting the barotropic index will be sufficient to solve the system.

Similar reasoning applies to our case, which adds hyperfluids. If we have a particular matter lagrangian where the matter fields explicitly couple to the non-Levi-Civita components of connection (or equivalently to torsion and/or nonmetricity), we can directly calculate the components of the hypermomentum tensor and deduce the relations between the spin and shear components among themselves or to energy density and pressure. Again, in principle, many fundamental Lagrangians can give the same relations. In this paper, we do not want to fix the fundamental Lagrangian but aim to study how the inclusion of hypermomentum affects the cosmological evolution in general under some reasonable assumptions, and in reverse, what properties should the hyperfluid have in order to engender certain cosmological evolution. For example, we may impose that spin or shear are fixed by the value of the energy density $\rho$, or that the hypermomentum index \eqref{eq: w_h definition} is constant, or that the energy-momentum and hypermomentum quantities obey separate continuity equations, i.e.\ the left and right sides of \eqref{eq: continuity eq} vanish independently, or that the expansion of the universe obeys certain form, i.e.\ fixing $H$ or $\w_{\mathrm{eff}}$.

\section{Fluid without hypermomentum}
\label{GR}

Before proceeding to study the models with hyperfluid, it is first instructive to recall the basic cosmology in general relativity. 
This little exercise will help to facilitate later comparison with new effects that arise from hypermomentum.

With vanishing hypermomentum and barotropic matter fluid \eqref{eq: w definition}, the equations \eqref{eq: FLRW equations general}
reduce to 
\begin{subequations}
\label{eq: FLRW equations general relativity}
\begin{align}
\label{eq: FR1 GR}
    3 H^2 &= \kappa \rho \,, \\
\label{eq: FR2 GR}
    2 \dot{H} + 3 H^2 &= - \kappa \w \rho \,, \\
\label{eq: continuity eq GR}
    \dot{\rho} + 3 H (\rho + p ) &= 0 \,.
\end{align}
\end{subequations}
To find the time dependence of matter-energy density, we can express the Hubble function from \eqref{eq: FR1 GR},
\begin{align}
\label{eq: FR1 H GR}
    H &= \pm \sqrt{\frac{\kappa \rho}{3}} \,,
\end{align}
and substitute that into \eqref{eq: continuity eq GR},
\begin{align}
\label{eq: density GR eq}
    \dot{\rho} \pm \sqrt{3 \kappa} (1+\w_\rho) \rho^{3/2} &=0 \,.
\end{align}
Here we have denoted $\w=\w_\rho$ to stress that the barotropic index in this equation determines how fast will 
the energy density dilute as the universe expands. For later purposes, we will call $\w_\rho$ a density 
index since, in general, it will differ from the barotropic index $\w$. 
The last equation can be integrated to
\begin{align}
    \rho(t) = \frac{4}{ \left(C_1 \mp \sqrt{3 \kappa} (1 + \w_\rho) \, t  \right)^2} \,.
\end{align}
We can fix the integration constant $C_1$ by asking that at some arbitrary moment $t_0$ (which may coincide with the 
present moment), the energy density is given by $\rho(t_0)=\rho_0$. This yields
\begin{align}
    C_1 &= \frac{\pm \sqrt{3\kappa \rho_0} (1+\w_\rho) t_0 - 2}{\sqrt{\rho_0}} \,.
\end{align}
To be precise, the integration constant has two solutions for expanding and two for contracting universes, but 
only one of those is consistent with positive time direction, i.e.\ not increasing energy density of non-phantom 
($\w\geq -1 $) matter in expanding universe and not decreasing energy density of non-phantom ($\w\geq -1 $) 
matter in a contracting universe. The final expression then reads
\begin{align}
\label{eq: rho(t) GR}
    \rho(t) &= \frac{\rho_0}{\left(1 \pm \frac{\sqrt{3\kappa \rho_0}}{2} (1+\w_\rho) (t-t_0) \right)^2} \,,
\end{align}
where the $\pm$ signs correspond to those of Eq.\ \eqref{eq: FR1 H GR}.

In a similar vein, we can solve for the Hubble function by substituting $\rho$ from \eqref{eq: FR1 GR} into \eqref{eq: FR2 GR} 
to obtain
\begin{align}\label{wefforigin}
    2 \dot{H} + 3 (1+\weff) H^2 &= 0 \,.
\end{align}
Here we have denoted $\w=\weff$ to stress that in this context, the barotropic index controls the universe expansion. 
The equation integrates to
\begin{align}
    H &= \frac{2}{C_2 + 3 (1+\weff) t} \,,
\end{align}
whereby the integration constant $C_2$ can be fixed by the requirement that at certain moment $t_0$ the Hubble parameter 
is $H(t_0)=H_0$, which implies
\begin{align}
    C_2 &= -\frac{3 (1+\weff) H_0 t_0 -2 }{H_0} \,.
\end{align}
Thus, the final result is
\begin{align}
\label{eq: H GR)}
    H &= \frac{H_0}{1+\frac{3 H_0}{2}(1+\weff)(t-t_0)} \,.
\end{align}

Of course, the expressions \eqref{eq: rho(t) GR} and \eqref{eq: H GR)} agree with the Friedmann equation \eqref{eq: FR1 GR} when 
\begin{align}
\label{eq: H0 GR}
    H_0 &= \pm \sqrt{\frac{\kappa \rho_0}{3}}
\end{align}
and $\w = \w_\rho = \weff$. Clearly, we could have obtained \eqref{eq: H GR)} just algebraically by substituting 
\eqref{eq: rho(t) GR} into \eqref{eq: FR1 GR} or in reverse obtained \eqref{eq: rho(t) GR} by substituting \eqref{eq: H GR)} 
into \eqref{eq: FR1 GR}. We will see later that in many cases, the inclusion of hypermomentum means the indices describing 
the dilution of matter, $\w_\rho$, and the expansion of the universe, $\weff$, can be different from each other, 
as well as from the index $\w$, which describes the microscopic properties of matter.

\section{Single component hypermomentum}
\label{sec: 5 Single component hypermomentum}

In order to get a better understanding of how the hypermomentum components can affect the cosmological evolution, 
we will first consider the cases of $\sigma$, $\Sigma_1$, and $\Sigma_2$ separately.

\subsection{Only spin}
\label{subsec: only spin}

Taking $\Sigma_1=\Sigma_2=0$, the equations \eqref{eq: FR1} and \eqref{eq: continuity eq} with only spin reduce to
\begin{subequations}
\label{eq: spin only}
\begin{align}
\label{eq: FR1 spin only}
    3 H^{2} &= \kappa\rho + 3 H \kappa \sigma - \frac{3 \kappa^{2} \sigma^{2}}{4} \,, \\
    2 \dot{H} + 3 H^{2} &= - \kappa p + 2 H \kappa \sigma  + \dot{\sigma} \kappa - \frac{\kappa^{2} \sigma^{2}}{4} \,, \\
\label{eq: continuity spin only}
    \dot{\rho} + 3 H \left( \rho + p \right) &= \frac{\kappa \sigma}{2} \left( \rho + 3 p \right) \,,
\end{align}
\end{subequations}
and the expression \eqref{eq: w_h definition} states
\begin{align}
\label{eq: w_h spin only}
    \w_h &= - \frac{8 H \sigma + 4 \dot{\sigma} - \kappa \sigma^{2}}{3 \sigma \left(4 H - \kappa \sigma\right)} \,.
\end{align}
We can express from \eqref{eq: FR1 spin only} the Hubble function
\begin{align}
\label{eq: H spin only}
    H &= \frac{\kappa \sigma}{2} \pm \sqrt{\frac{\kappa \rho}{3}} \,,
\end{align}
Hence, the solutions come in two branches, as in general relativity.

Now, in \eqref{eq: spin only}, we have four unknowns $\rho(t),p(t), \sigma(t), H(t)$, and only two independent equations. In order to solve the system, we have to impose two constraints. For example, we may impose that one or both barotropic indexes \eqref{eq: w definition}, \eqref{eq: w_h definition} are constants, or that the spin is proportional to the energy density, or that the fluid and hyperfluid quantities obey separate continuity equations, i.e.\ the left and right sides of \eqref{eq: continuity spin only} vanish independently, or that the expansion of the universe obeys certain form, i.e.\ fixing $H$ or $\weff$.

\subsubsection{Assuming that spin is proportional to energy density}
\label{subsubsec: spin proportional energy density}

Comparing the first and last terms on the r.h.s. \ of Eq. \eqref{eq: FR1 spin only}, we notice that dimensionally 
the $\kappa^2 \sigma^2$ compares to $\kappa \rho$. From that, we may assume a direct proportionality 
\begin{align}
\label{eq: sigma(rho) assume}
    \sigma &= b \sqrt{\frac{3 \rho}{\kappa}} \,,
\end{align}
where $b$ is a dimensionless constant. Now, also assuming that $\w$ is constant leaves only $\rho$ and $H$ as independent variables, and we can fully solve the system. Substituting $H$ from \eqref{eq: H spin only} into \eqref{eq: continuity spin only} gives
\begin{align}
\label{eq: sigma2 rho}
    \dot{\rho} \pm \sqrt{3 \kappa} \left(1+\w\pm b\right)\rho^{3/2} =& 0 \,,
\end{align}
where the $\pm$ signs represent the two branches of $H$ in \eqref{eq: H spin only}. Comparing \eqref{eq: sigma2 rho} with Eq.\ \eqref{eq: density GR eq} we see that the only influence of the spin is to effectively modify the barotropic index of the matter, giving the time dependence of $\rho(t)$ as in \eqref{eq: rho(t) GR} with
\begin{align}
\label{eq: w_rho spin prop to rho}
    \w_\rho &= \w \pm b \,.
\end{align}

From \eqref{eq: H spin only} and \eqref{eq: sigma(rho) assume} it is also straightforward to compute the expansion rate of the universe, which follows Eq.\ \eqref{eq: H GR)} with the effective barotropic index given by
\begin{align}
\label{eq: weff spin prop to rho}
    \weff &= \frac{2 \w \mp b}{2 \pm 3b} \,.
\end{align}
This shows that in contrast to normal fluids where the effect on the expansion and the dilution of energy density is characterized by a single number $\w$, the hyperfluid cosmology behaves differently as the effect on expansion and the dilution of energy density are set by different numbers \eqref{eq: weff spin prop to rho} and \eqref{eq: w_rho spin prop to rho}, respectively. Note that for the specific value of the spin coupling
\beq
b=\mp (1+\w)
\eeq
the two indexes become identical as in the GR case. The Hubble parameter at $t_0$ can be simply found from \eqref{eq: H spin only} with the change of variables \eqref{eq: sigma(rho) assume}
\begin{equation}
\label{eq: H0 spin prop to rho}
    H_0 =\frac{(3 b \pm 2) }{2} \sqrt{ \frac{\kappa \rho_0 }{3}} \, .
\end{equation}
In addition, we may also compute the barotropic index $\w_h$ of the spin part separately by substituting $\dot{\rho}$ from \eqref{eq: sigma2 rho} and $H$ from \eqref{eq: H spin only} into Eq. \eqref{eq: w_h spin only}, then $\rho$ cancels and we obtain
\begin{align}
    \w_h =& \pm \frac{6 \w - 2 \mp 3 b }{3 \left( \pm 4 + 3 b \right)} \,,
\end{align}
consistent with our definitions \eqref{eq: w definition}, \eqref{eq: w_eff definition general}.
In retrospect, the latter is nothing else than equating Eqs.\ \eqref{eq: sigma(rho)} and \eqref{eq: sigma(rho) assume}, and expressing $\w_h$. In any case, these results offer new possibilities to model cosmological dynamics.
For instance, in the positive $H$ branch if $\w=1/3$ (radiation) and $b=-1/3$ then both $\rho$ and $\sigma^2$ dilute as fast as dust, but in the second Friedmann equation the spin exerts extra negative pressure with $\w_h=-1/15$ while the overall pressure corresponds to $\weff=1/9$.
For later comparisons let us also expand \eqref{eq: w_rho spin prop to rho}, \eqref{eq: weff spin prop to rho},  
\eqref{eq: H0 spin prop to rho} for small $b$ to get
\begin{subequations}
\label{eq: w_rho, w_eff, H_0 small b}
    \begin{align}
        \w_\rho &\approx \w \pm b \,, \\
        \weff &\approx \w \mp \frac{(3\w+1)b}{2} \,, \\
        H_0 &\approx \left( \pm 1 + \frac{3b}{2} \right) \sqrt{\frac{\kappa \rho_0}{3}} \,.
    \end{align}
\end{subequations}
It is obvious that in the limit of vanishing spin ($b = 0$) the indices coincide, $\w_\rho=\weff=\w$, 
and the Hubble constant reduces to the value in general relativity, Eq.\ \eqref{eq: H0 GR}.

\subsubsection{Assuming constant hypermomentum index}
\label{subsec: spin both w constant}

Next, we will assume that $\w$ and $\w_h$ are constant. This is surely a strong hypothesis for  $\w_h$ considering its definition \eqref{eq: w_h definition} combined with \eqref{rhohph}. However, we saw in the previous section \ref{subsubsec: spin proportional energy density} that a simple hypothesis about the relation between spin and matter density does indeed give a constant $\w_h$.

Hence, in the case when $\w$ and $\w_h$ are both constant, we can use $\eqref{eq: H spin only}$ inside \eqref{eq: FR1 spin only} and \eqref{eq: continuity spin only} so to obtain
\begin{subequations}
\label{eq: spin only r}
\begin{align}
\label{eq: FR1 spin only r}
    \dot{\rho} &=  - \frac{\left( \pm 4 \sqrt{3 \kappa \rho} \left(3 \w_h + 2\right) + 9 \kappa \left(\w_h + 1\right) \sigma{\left(\rho \right)}\right) \sigma{\left(\rho \right)}}{12 \frac{d \sigma}{d \rho} } \,, \\
\label{eq: continuity spin only r}
    \dot{\rho} &= - \frac{\kappa \rho \left(9 \w \w_h + 9 \w - 3 \w_h + 1\right) \sigma^{2}{\left(\rho \right)}}{4 \left(3 \rho \left(\w + 1\right) \frac{d \sigma}{d \rho}  - \left(3 \w_h + 2\right) \sigma{\left(\rho \right)}\right)}
\end{align}
\end{subequations}
by expecting that the spin is a function of matter density, $\sigma(\rho)$. 
Equating the right hand sides of \eqref{eq: FR1 spin only r} and \eqref{eq: continuity spin only r}, we can express
\begin{align}
\label{eq: sigma(rho) equation}
    \frac{\dd \sigma}{\dd \rho} &= \frac{\left(4 \sqrt{3 \kappa \rho} \left(3 \w_h + 2\right) \pm 9 \kappa \left(\w_h + 1\right) \sigma{\left(\rho \right)}\right) \sigma{\left(\rho \right)}}{12 \rho \left(\sqrt{3 \kappa \rho} \left(\w + 1\right) \pm \kappa  \sigma{\left(\rho \right)}\right)} \,,
\end{align}
which is possible to integrate to
\begin{align}\label{intrho}
-\sigma(\rho)^{6(1+3\w_h)(1+\w)} \rho^{-3(1+\w_h)(1-3\w+6\w_h)} 
\left(2\sqrt{3 \kappa \rho} (1-3\w+6\w_h) \pm 3 \kappa (1+3\w_h) \sigma(\rho) \right)^{-2 + 6\w_h-18 \w_h \w-18\w} &= C \,.
\end{align}
It is reasonable to assume that spin vanishes when matter density is zero, $\sigma(0)=0$, which fixes the integration constant $C=0$, and yields for general $\w$ and $\w_h$
\begin{align}
\label{eq: sigma(rho)}
    \sigma(\rho) &= \pm \frac{2 \left(3 \w - 6 \w_h - 1\right)}{3 \w_h + 1}\sqrt{\frac{\rho}{3 \kappa}}  \,.
\end{align}
Substituting the expression \eqref{eq: sigma(rho)} into either of the equations \eqref{eq: spin only r} we obtain
\begin{align}
\label{eq: rho eq in sigma}
    \dot{\rho} \pm \sqrt{3 \kappa} (1+\w_\rho) \rho^{3/2} &= 0
\end{align}
where 
\begin{align}
\label{eq: 1+w_rho with spin}
    1+\w_\rho &= \frac{9 \w \w_h + 9 \w - 3 \w_h + 1}{3 \left(3 \w_h + 1\right)}
\end{align}
is a constant. The equation \eqref{eq: rho eq in sigma} is structurally identical to the corresponding equation in general relativity \eqref{eq: density GR eq} with the only difference in how the quantity $\w_\rho$ depends on the microscopic indices $\w$ and $\w_h$. It means the solution \eqref{eq: rho(t) GR} is also valid here with only $\w_\rho$ now given by \eqref{eq: 1+w_rho with spin}. Hence, we see that the inclusion of spin can significantly affect how fast the energy density dilutes when the universe expands. Moreover, let us note that the assumption of constant $\w_h$ does not allow us to connect straightforwardly to the GR limit as instead does the \eqref{eq: w_rho, w_eff, H_0 small b} with $b=0$ in previous section.

Now \eqref{eq: sigma(rho)} immediately gives the time dependence of the spin, and \eqref{eq: H spin only} establishes the expansion (Hubble function) structurally in the same form as in general relativity, \eqref{eq: H GR)}, only
\begin{align}
    H_0 &= \pm \frac{\sqrt{3 \kappa \rho_0} (\w-\w_h)}{(1+3\w_h)}
\end{align}
and
\begin{align}
\label{eq: 1+w_eff with spin}
    1+\weff &= \frac{9 \w \w_h + 9 \w - 3 \w_h + 1}{9 \left(\w - \w_h\right)} \,,
\end{align}
thus completely solving the system. This shows that the inclusion of spin necessarily affects the expansion of the universe as well. Moreover, comparing the expressions \eqref{eq: 1+w_rho with spin} and \eqref{eq: 1+w_eff with spin}, we see that the effective barotropic index $\w_\rho$ which describes the dilution of matter density and the index $\weff$ which describes the expansion of the universe differ from each other.
However, when $\weff=-1$ we also get $\w_\rho=-1$, thus de Sitter expansion is consistent with constant matter density. The latter happens for $\w_h=\frac{1+9\w}{3(1-3\w)}$, e.g.\ for dust matter ($\w=0$) the index for spin should be $\w_h=1/3$. Again we can see how the assumption of $\w_h$ constant does not give us the same GR limit as instead does the \eqref{eq: w_rho, w_eff, H_0 small b} with $b=0$ in the previous section.

In the calculation above, arbitrary constants $\w$ and $\w_h$ were assumed. In certain particular cases, the results would be different. For instance, if $\w = \w_h$, then 
\begin{align}
\label{eq: sigma w=w_h spin}
    \sigma &= \mp 2\sqrt{\frac{\rho}{3\kappa}}
\end{align}
and by \eqref{eq: H spin only} the contributions of energy and spin cancel each other in the Friedmann equation, thus $H$ is identically zero. Curiously, this does not mean that the energy density and spin are constant, the former still evolves as Eq.\ \eqref{eq: rho(t) GR} with 
\begin{align}
    1+\w_\rho = \frac{1}{3} + \w_h \,
\end{align}
and the latter can then be deduced from \eqref{eq: sigma w=w_h spin}.
Similarly, if $\w_h=-1/3$ and $\w=-1$, the equation \eqref{eq: sigma(rho) equation} simplifies considerably and again integrates to the same expression as \eqref{eq: sigma w=w_h spin}.
In the particular cases $\w_h=-1/3$ (with $\w \neq -1$), and $\w_h=-\frac{1}{6}+\frac{\w}{2}$ the nature of the equation \eqref{eq: sigma(rho) equation} changes. The analytic integration is still possible, but the expressions are long and contain special functions, and we will skip the details of these solutions in the paper.

\subsubsection{Assuming separate conservation laws for energy and spin}
\label{subsec: separate conservation spin}

If $\w$ and $\w_h$ are both constant as in the considerations above, the fluid and hyperfluid components are not separately conserved, i.e., the left and right side of \eqref{eq: continuity spin only} are not zero separately. However, it may make sense to assume a separate continuity equation holding for the energy density at least, e.g.,  that the density of dust matter particles would change inversely proportional to the changing volume, independent of whether the particles carry hypermomentum or not. Thus, if we assume matter conservation, implying that both sides of \eqref{eq: continuity spin only} vanish separately, then from the r.h.s. we see immediately that
\begin{align}
    \w=-1/3 \,,
\end{align}
which in the l.h.s.\ implies $\rho\sim a^{-2}$, i.e.\ the matter can be neither radiation nor dust. To solve the system fully, we need to make yet another assumption. Taking $\w_h$ to be constant, we just get a particular subcase of the result obtained before. If we do not assume anything about $\w_h$, we could combine \eqref{eq: H spin only} and \eqref{eq: continuity spin only} to 
\begin{align}\label{sepcons}
    \dot{\rho} \pm \frac{2}{3} \sqrt{ 3 \kappa} \rho^{3/2} + \kappa \sigma \rho &=0 \,.
\end{align}

We may solve \eqref{sepcons} by assuming some spin equation of state $\sigma(\rho)$. In particular, an ansatz 
like \eqref{eq: sigma(rho) assume} would take us back to the case considered in the previous subsection.
Another option to solve  \eqref{sepcons} would be to demand the expansion to obey a certain rule, 
e.g.\ to be de Sitter with constant $H=H_*$, whereby the l.h.s.\ of \eqref{eq: continuity spin only} and 
\eqref{eq: H spin only} and  are satisfied by 
\begin{align}
\label{eq: spin rho H separate conservation}
    \rho &= \rho_0 e^{-2 H_* (t-t_0)} \,, \qquad \sigma = \frac{2 H_*}{\kappa} \mp 2 \sqrt{\frac{\rho_0}{3 \kappa}}  
    e^{-H_*(t-t_0)} \,.
\end{align}
Here the spin must sustain a constant part of its density to maintain such expansion, which remains nonzero even when 
the matter density vanishes. Hence, it is only a formal solution hard to realize by some physical substance.

\subsubsection{Assuming de Sitter expansion}
\label{subsubsec: de Sitter spin only}

In general, if we impose de Sitter expansion $H=H_*$, i.e.\ the law \eqref{eq: H GR)} with 
\begin{align}
    \weff=-1 \,,
\end{align}
and also assume an arbitrary constant equation of state $\w$ for the matter fluid, then from \eqref{eq: H spin only} it follows that 
\begin{align}
\label{sol:deSitter:sigma}
    \sigma &= \frac{2H_*}{\kappa} \mp 2\sqrt{\frac{\rho}{3 \kappa}} \,.
\end{align}
Here, the spin must carry a constant contribution which is independent of the matter density and is hard to justify physically. This is quite obvious, as in order to have constant $H$ in \eqref{eq: H spin only}, either the matter density or spin must have a constant term while the time-dependent parts of density and spin cancel, analogously to the solution \eqref{eq: spin rho H separate conservation} encountered before.

Without imposing a further assumption that the energy density and spin are separately conserved, we can substitute $H_*$ and \eqref{sol:deSitter:sigma} into \eqref{eq: continuity spin only} to get
\begin{align}
    \dot{\rho} \pm  \sqrt{\frac{\kappa}{3}} (1 + 3 \w) \, \rho^{{3}/{2}} + 2 H_{*} \rho = 0\,.
\end{align}
This equation has an extra term compared to \eqref{eq: density GR eq}, but it can still be integrated. Fixing the integration constant by setting $\rho(t_0)=\rho_0$ the final expression is
\begin{align}\label{rhodesitter}
    \rho(t) &= \frac{\rho_0}{e^{2H_*(t-t_0)} \left(1 \pm \sqrt{\frac{\kappa \rho_0}{3}} \frac{(1+3\w)}{2 H_* } \left(1-e^{-H_*(t-t_0)}\right)\right)^2} \,.
\end{align}
The energy density decays exponentially, and the constant part of the spin \eqref{sol:deSitter:sigma} maintains the de Sitter acceleration of the universe.
In \eqref{rhodesitter} we have chosen the sign of the integration constant as in Eq.\ \eqref{eq: rho(t) GR}.

\subsubsection{Weyssenhoff fluid inspired model}
In the context of Einstein--Cartan theory which has only torsion and no nonmetricity, a popular model for the source of torsion is the  Weyssenhoff fluid 
\cite{Weyssenhoff:1947iua} (see also \cite{Ray:1982qs, Ray:1982qr, Ray:1982qq, Kopczynski:1986ep, Kopczynski:1990af}), 
whereby the spin part of hypermomentum is linked to the quantum mechanical 
intrinsic spin density of a fermionic fluid. In an unpolarized case the quantum fluctuations 
of the spin orientation are random, and the average of the spin density will vanish, but the terms that are 
quadratic in spin have a nonvanishing average which can be related to the energy density as 
\cite{Gasperini:1986mv, Luz:2018poo,  Medina:2018rnl, Izaurieta:2020xpk, Akhshabi:2023xan}
\begin{align}
\label{eq: Weyssenhoff s2}
    s^2 &= \mathcal{B}_\w \, \rho^{\frac{2}{1+\w}} \,,
\end{align}
where $\mathcal{B}_\w$ is a dimensionful constant depending on the barotropic index $\w$.
However, the construction of the Weyssenhoff fluid
does not satisfy the cosmological symmetry \cite{tsamparlis1979cosmological, Boehmer:2006gd} and 
thus does not fall under the purview of the present study. Despite that, we can still adapt the relation 
\eqref{eq: Weyssenhoff s2} as a possible equation state to solve the cosmological field equations.

Firstly, we can just assume $\sigma =  B_\w \, \rho^{\frac{1}{1+\w}}$, and substitute $H$ from the first Friedmann 
equation \eqref{eq: FR1 spin only} into the continuity equation \eqref{eq: continuity spin only}. 
The result is the usual general relativity expression augmented by an extra correction term,
\begin{align}
\label{eq: sigma rho power}
    \dot{\rho} &= \mp \sqrt{3 \kappa} (1+\w) \rho^{\frac{3}{2}} + \kappa B_\w \rho^{\frac{\w+2}{\w+1}} \,.
\end{align}
In the stiff matter ($\w=1$) case it is obvious that the density obeys the general relativity rule 
\eqref{eq: density GR eq} with the following substitution
\begin{equation}
    1 + \w_{\rho} = 2 \mp \sqrt{\frac{\kappa}{3}} B_{\w} \,.
\end{equation} 
For other types of matter we can separate the variables in \eqref{eq: sigma rho power} as
\begin{align}
\label{eq: sigma rho power separate}
    \frac{\dd\rho}{\rho^{3/2}\left(1\mp \sqrt{\frac{\kappa}{3}} \frac{B_\w \rho^{\frac{1-\w}{2(1+\w)}}}{1+\w}\right)} &= \mp \sqrt{3\kappa}(1+\w) \dd t \,.
\end{align}
It is easier to investigate this for specific values of $\w$. For instance, in the case of dust matter ($\w=0$) Eq.\ \eqref{eq: sigma rho power separate} integrates to
\begin{align}
\label{sigma:weyssenhoff:sol:rho}
\mp 2 B_\w \sqrt{\frac{\kappa}{3}} \ln{\frac{\frac{1}{\sqrt{\rho}} \mp B_\w \sqrt{\frac{\kappa}{3}}}{\frac{1}{\sqrt{\rho_0}} \mp B_\w \sqrt{\frac{\kappa}{3}}}-\frac{2}{\sqrt{\rho}}+\frac{2}{\sqrt{\rho_0}}= \mp \sqrt{3\kappa} (t-t_0)} \,.
\end{align}
Approximating here $\ln x \approx x-1$ (valid for $0 < x \leq 2$, i.e.\ assuming that $\rho(t)$ is not too different from the reference value $\rho_0=\rho(t_0)$) allows to express
\begin{align}
    \rho(t) &\approx \frac{\rho_0}{\left(1 \pm \frac{\sqrt{3 \kappa \rho_0}}{2} \left(1 \mp B_\w\sqrt{\frac{\kappa \rho_0}{3}}\right) (t-t_0)\right)^2} \,.
\end{align}
A comparison with \eqref{eq: sigma rho power} shows that the dust matter density index is 
\begin{align}
    1 + \w_{\rho} &\approx 1 \mp \sqrt{\frac{\kappa\rho_0}{3}} B_{\w} \,,
\end{align} 
i.e.\ it is approximately zero as expected for dust matter if the second term in the r.h.s.\ of \eqref{eq: sigma rho power} is negligible, and starts to deviate from zero when that term starts to have influence in \eqref{eq: sigma rho power}. Interestingly, the density index $\w_\rho$ here depends on the reference value $\rho_0$, which means that different solutions in the model are characterized by different values of $\w_\rho$. The cases of other $\w$ can be also integrated and treated in an analogous manner.

Alternatively, we may assume $\rho_h=\mathfrak{B}_\w \, \rho^{\frac{1}{1+\w}}$ which brings the first Friedmann equation into the form studied in Refs.\ \cite{Medina:2018rnl}. However, in order to solve our system of equations \eqref{eq: spin only} we need to make yet another assumption, presumably postulating an equation of state that relates $\rho_h$ and $p_h$. Such step takes us generally back to subsection \ref{subsec: spin both w constant} where constant $\w_h$ was considered. Let us just remark that in this scheme the bouncing universe solution of Ref.\ \cite{Brechet:2007cj, Medina:2018rnl} corresponds to $\w_h=1$, $\rho_h<0$, while the spin fluid alternative to inflation corresponds to $\w_h=-1/3$, $\rho_h>0$ \cite{Gasperini:1986mv, Medina:2018rnl}.

\subsection{Only the first shear component}

Taking $\sigma=\Sigma_2=0$, the equations \eqref{eq: FR1} and \eqref{eq: continuity eq} contain only the first 
shear component and reduce to
\begin{subequations}
\label{eq: Sigma1 only}
\begin{align}
\label{eq: FR1 Sigma1 only}
    3 H^{2} &= \kappa\rho + 3 \kappa H \Sigma_1 \,, \\
\label{eq: FR2 Sigma1 only}
    2 \dot{H} + 3 H^{2} &= - \kappa p - \kappa H \Sigma_1 \,, \\
\label{eq: continuity Sigma1 only}
    \dot{\rho} + 3 H \left( \rho + p \right) &= -3 \left( H \dot{\Sigma}_1 + \dot{H} \Sigma_1 + 4 H^2 \Sigma_1 \right)
\end{align}
\end{subequations}
while the expression \eqref{eq: w_h definition} turns out to be
\begin{align}
\label{eq: w_h Sigma1 only}
    \w_h &= \frac{1}{3} \,.
\end{align}
This means that in the context of cosmology, the first shear component exerts a similar pressure to radiation. 
From \eqref{eq: FR1 Sigma1 only} we can express the Hubble function as 
\begin{equation}
\label{eq: H Sigma1 only}
H=\frac{ \kappa \, \Sigma_1 }{2} \pm \sqrt{\frac{\kappa \rho}{3} + \frac{\kappa^2 \Sigma_1^2}{4}} \,.
\end{equation}
where the $\pm$ sign corresponds to expanding and contracting universes, respectively. If $\frac{\kappa \Sigma_1^2}{\rho} \ll 1$ then we can expand the square root and approximate
\begin{align}
    H &\approx \frac{ \kappa \, \Sigma_1 }{2} \pm \sqrt{\frac{\kappa \rho}{3}} \,.
\end{align}

\subsubsection{Assuming that shear is proportional to energy density}
\label{subsubsec: shear1 proportional to energy density}

Like in Sec.\ \ref{subsubsec: spin proportional energy density} we may invoke the dimensionality argument and assume that
\begin{align}
    \Sigma_1 &= b_1 \sqrt{\frac{\rho}{3 \kappa}} \,.
\end{align}
Then it is possible to substitute $\dot{H}$ from \eqref{eq: FR2 Sigma1 only} and $H$ from \eqref{eq: H Sigma1 only} into \eqref{eq: continuity Sigma1 only} to find
\begin{align}\label{dotrho}
    \dot{\rho} \pm \frac{2 \sqrt{3 \kappa} \left( \pm 2 b_{1} \left(b_{1}^{2} + 4\right) + \sqrt{b_{1}^{2} + 4} \left(3 \w + 2 b_{1}^{2} + 3\right)\right) \rho^{3/2}}{3 \left(b_{1}^{2} \pm b_{1} \sqrt{b_{1}^{2} + 4} + 4\right)} &= 0 \,.
\end{align}
It is nice to witness that the equation \eqref{dotrho} has the same shape as \eqref{eq: density GR eq} in GR, only now
\begin{equation}
\label{eq: wrho Sigma1 b1}
    \w_\rho=\frac{\pm 4 b_1-3}{3}+\frac{2 (\w+1)}{\sqrt{b_1^2+4} \pm b_1} \,.
\end{equation}
Due to the congruence of the equations \eqref{eq: density GR eq} and \eqref{dotrho} the solution $\rho(t)$ is also structurally the same, \eqref{eq: rho(t) GR}, but now with $\w_\rho$ given by the expression above.
From \eqref{eq: H Sigma1 only} we can also immediately determine $H$ which follows 
\eqref{eq: H GR)}, only the constants get a modification that depends on the parameter $b_1$,
\begin{subequations}
\label{eq: weff H0 b1}
\begin{align}
    H_0 &= \frac{b_1 \pm \sqrt{b_1^2+4}}{2}\sqrt{\frac{\kappa \rho_0}{3}} \,, \\
\label{eq: weff Sigma1 b1}
    \weff &= \frac{1}{3} + \frac{6 \w -2}{3 \left( b_1^2 \pm b_1 \sqrt{b_1^2 +4} +2 \right)} \,.
\end{align}
\end{subequations}
In the limit where the shear component is small in comparison to the energy density, i.e.\ $b_1 \ll 1$, the expressions \eqref{eq: wrho Sigma1 b1} and \eqref{eq: weff H0 b1} approximate to
\begin{subequations}
\label{eq: S1 b1 small}
\begin{align}
    \w_\rho & \approx \w \mp \frac{(3 \w -5) b_1}{6} \,, \\
    \weff &\approx \w \mp \frac{(3 \w -1) b_1}{3} \,, \\
    H_0 & \approx \left( \pm 1 +\frac{b_1}{2} \right) \sqrt{\frac{\kappa \rho_0}{3}}
    \,.
\end{align}
\end{subequations}
Thus a hyperfluid with extra shear component $\Sigma_1$ that is proportional to its energy density $\rho$ will run a cosmology rather similar to a normal fluid, only the evolution of the energy density and spacetime expansion are now ruled by modified indices.

\subsubsection{Assuming separate conservation laws for energy and shear first component}
\label{subsubsec: separate conservation laws shear1}

Since $\w_h$ is fixed, it does not provide an extra condition that can be used to reduce the number of variables. However, we can assume that both sides of Eq.\ \eqref{eq: continuity Sigma1 only} hold separately, and use the same approach as in Sec.\ \ref{subsec: spin both w constant}. Namely, if we substitute $\dot{H}$ from \eqref{eq: FR2 Sigma1 only} and $H$ from \eqref{eq: H Sigma1 only} into the right and left sides of \eqref{eq: continuity Sigma1 only} we can express 
\begin{subequations}
\label{rhodotsigma1}
\begin{align}
    \dot{\rho} =& \pm \frac{\left( \kappa \rho \left(3 \w - 5\right) - 2 \kappa \left( \pm \sqrt{3 \kappa} \sqrt{3 \kappa \Sigma_{1}^{2}{\left(\rho \right)} + 4 \rho} + 3 \kappa \Sigma_{1}{\left(\rho \right)}\right) \Sigma_{1}{\left(\rho \right)}\right) \Sigma_{1}{\left(\rho \right)}}{\left(\sqrt{3 \kappa} \sqrt{3 \kappa \Sigma_{1}^{2}{\left(\rho \right)} + 4 \rho} \pm 3 \kappa \Sigma_{1}{\left(\rho \right)}\right) \frac{\dd \Sigma_1}{\dd \rho} } \,, \\
\label{rhodotsigma1b}
    \dot{\rho} =& - \frac{\rho \left(\w + 1\right) \left( \pm \sqrt{3 \kappa} \sqrt{3 \kappa \Sigma_{1}^{2}{\left(\rho \right)} + 4 \rho} + 3 \kappa \Sigma_{1}{\left(\rho \right)}\right)}{2} \,,
\end{align}
\end{subequations}
where it was assumed that $\Sigma_1$ is a function of $\rho$. Combining these two equations gives

\begin{equation}\label{eq: Sigma1(rho) eq conservation case}
    \frac{\dd \Sigma_1}{\dd \rho} =\frac{2 \Sigma_{1}}{3(1+\w)\rho}+\frac{(1-3 \w)}{18(1+\w)}\frac{\kappa\Sigma_{1}}{\left(\frac{\kappa \Sigma_{1}}{2} \pm \sqrt{ \frac{\kappa \rho}{3} + \frac{\kappa^2 \Sigma_{1}^{2}}{4} } \right)^2}
\end{equation}

If $\w=1/3$ we get $\Sigma_1 \propto \sqrt{\rho}$, and situations like this will be considered in more detail in Sec.\ \ref{subsubsec: shear1 proportional to energy density}. For general $\w$ the equation can be solved by a change of variables, introducing
\begin{align}
\label{separ}
    \Sigma_{1}(\rho) &=\sqrt{\frac{\rho}{ 3 \kappa}}f(\rho) \,.
\end{align}
with the dimensionless function $f(\rho)$.
Then the Eq.\ \eqref{eq: Sigma1(rho) eq conservation case} takes a form
\beq
\rho \frac{\dd f}{\dd \rho}=\frac{(1-3\w)}{6(1+\w)}f\left[ 1+ \frac{1}{\Big(\frac{f}{2} \pm \sqrt{1+\Big(\frac{f}{2}\Big)^2} \Big)^{2}} \right] \label{separdiff}
\eeq
Notice the common factor that has formed on the right-hand side, which allows one to find a complete exact solution as we now show. Indeed, we have reduced the problem of solving a differential equation to that of computing a simple integral. Performing the integration, we obtain\footnote{The integral can be easily evaluated by first the subsequent change of variables $f=2\sinh{\theta}$ and $u=e^{\theta}$, and keeping in mind that the argument of $\ln$ must be positive.}
\beq
\label{fsolution}
f(\rho)=\pm\frac{C\rho^{\frac{(1-3 \w)}{3(1+\w)}}}{\sqrt{1 \pm C\rho^{\frac{(1-3 \w)}{3(1+\w)}}} } \,,
\eeq
and therefore, the solution of \eqref{separdiff} for the shear reads
\beq
\label{S1solution}
\Sigma_{1}(\rho)=\pm \frac{C\rho^{\frac{(1-3 \w)}{3(1+\w)}}}{\sqrt{1\pm C\rho^{\frac{(1-3 \w)}{3(1+\w)}}} } \sqrt{\frac{\rho}{3 \kappa}} \,.
\eeq
For the initial condition $\Sigma_{1}(\rho_{0})=\Sigma_{1,0}$ the integration constant is fixed as
\beq
C = \pm \frac{3 \kappa}{2} \Sigma_{1,0}^2 \, \rho_0^{-\frac{4}{3 (\w+1)}} \left(1 + \sqrt{ 1 + \frac{4\rho_0}{3 \kappa \Sigma_{1,0}^2}}\right) \,.
\eeq
Substituting the above solution \eqref{S1solution} into the continuity equation for $\rho$ \eqref{rhodotsigma1b} gives 
\begin{align}
    \dot{\rho} \pm \sqrt{3\kappa} (1+\w) \rho^{3/2} \sqrt{1\pm C\rho^{\frac{1-3\w}{3(1+\w)}}} &=0 \,.
\end{align}
Here the second term takes the same form as in GR \eqref{eq: continuity eq GR} multiplied by a correction factor. For general $\w$ the equation above can be integrated in terms of hypergeometric functions, but the extraction of the final analytic expression of $\rho(t)$ is difficult. For instance, in the case $\w=0$, the result is a sixth-order polynomial equation for $\rho$.

If we assume $\frac{\kappa \Sigma^2_1}{\rho} \ll 1$ and develop the r.h.s.\ of \eqref{eq: Sigma1(rho) eq conservation case} into series, then the two leading terms are
\begin{align}\label{sigma1approx}
    \frac{\dd \Sigma_1}{\dd \rho} &\approx  \frac{(5-3\w) \Sigma_1(\rho)}{6 (1+\w)\rho} \pm \frac{\sqrt{3 \kappa}(3 \w -1) \Sigma_1^2(\rho)}{6 (1+\w) \rho^{3/2} } \,.
\end{align}
As solution of \eqref{sigma1approx} we get 
\begin{equation}\label{sigma1approxsol}
 \Sigma_1(\rho )= \pm \frac{2}{1- \left( 1 \mp \frac{2 }{\Sigma_{1,0}} \sqrt{\frac{\rho_0}{3 \kappa}}\right) \left(\frac{\rho }{\rho_0}\right)^{\frac{(3 \w-1)}{3 (\w+1)}}} \sqrt{\frac{\rho }{3 \kappa}} \,.
\end{equation}
For late universe with small $\rho$, the regime of $\kappa \Sigma^2_1 \ll \rho$ can be maintained if $-1 < \w < 1/3$. In the other case of $\w < -1$ or $\w > 1/3$, the second term in the denominator in \eqref{sigma1approxsol} becomes insignificant, and the shear component enters a regime of $\kappa \Sigma_1^2 \sim \rho$ that we are going to study in the next subsection.

\subsubsection{Assuming de Sitter expansion}

Let us assume de Sitter expansion $H=H_*$ and arbitrary constant equation of state $\w$ for the matter fluid. The main difference from the analogous spin case of Sec.\ \ref{subsubsec: de Sitter spin only} is that the first shear component $\Sigma_1$ appears in the Friedmann equation \eqref{eq: FR1 Sigma1 only} only in the friction term multiplied by $H$ and not by itself like $\rho$ or $\sigma^2$. The the equations \eqref{eq: Sigma1 only} combined imply
\begin{align}\label{shear1ds}
    \Sigma_1 &= -\frac{(1+\w) \rho}{4 H_*} \,,
\end{align}
i.e.\ the shear component must be negative for non-phantom $\w$. For the $\w=-1$, the shear part completely drops out, as the cosmological constant itself is compatible with de Sitter expansion.  Substituting \eqref{shear1ds} into \eqref{eq: FR1 Sigma1 only} yields
\begin{align}
    3 H_*^2 &= \frac{1-3\w}{4} \kappa \rho \,,
\end{align}
which means the energy density must be constant, $\rho=\rho_*$, and hence shear must be also constant, $\Sigma_1=\Sigma_{1,*}$. This is an interesting result, as we do not need the usual $p = -\rho$ relation, and even not necessarily negative pressure at all to maintain features that in GR are associated with a cosmological constant. Instead, we could just have a dust type ($\w=0$) hyperfluid carrying negative shear to get a de Sitter universe with exponential expansion of space and energy density that remains constant despite the expansion. The reason for that is that now the continuity equation is modified.
For a fixed value $H_{*}$ and reasonable thermodynamic properties of matter, $-1 \leq \w \leq 1$, we get a constraint for the density,
\beq
-1 \leq \frac{6 H_{*}^{2}}{\kappa \rho_{*}}\leq 2 \,.
\eeq
Let us note that things here are quite different from the pure spin where the de Sitter phase there was supported by non-constant energy density and spin see Eqns.\ (\ref{rhodesitter}), (\ref{sol:deSitter:sigma}).

\subsection{Only the second shear component}

Taking $\sigma=\Sigma_1=0$, the equations \eqref{eq: FR1} and \eqref{eq: continuity eq} reduce to
\begin{subequations}
\label{eq: Sigma2 only}
\begin{align}
\label{eq: FR1 Sigma2 only}
    12 H^{2} &= 4 \kappa \rho + 18 \kappa H \Sigma_{2} + 3 \kappa^{2} \Sigma_{2}^{2} + 6 \kappa \dot{\Sigma}_2   \,, \\
\label{eq: FR2 Sigma2 only}
    8 \dot{H} + 12 H^{2} &= - 4 \kappa p - 6 \kappa H \Sigma_{2} - 3 \kappa^{2} \Sigma_{2}^{2} - 2 \kappa \dot{\Sigma}_2  \,, \\
\label{eq: continuity Sigma2 only}
    \dot{\rho} + 3 H \left( \rho + p \right) &= - \frac{3 \ddot{\Sigma}_2}{2} - 18 H^{2} \Sigma_{2} - \frac{9 \kappa H \Sigma_{2}^{2} }{2} - \frac{21 H \dot{\Sigma}_2}{2} - \frac{9 \Sigma_{2} \dot{H}}{2} - \frac{3 \kappa \Sigma_{2} \dot{\Sigma}_2}{2} \,.
\end{align}
\end{subequations}
The expression \eqref{eq: w_h definition} turns out to be
\begin{equation}\label{eq: w_h Sigma2 only}
   \w_h = \frac{1}{3} + \frac{2 \, \kappa \, \Sigma_{2}^2}{ 3 \left(6 H \Sigma_{2}+ \kappa \, \Sigma_{2}^2 + 2 \, \dot{\Sigma}_{2} \right)} \,,
\end{equation}
while the expression of H from \eqref{eq: FR1 Sigma2 only} is
\begin{equation}
\label{H Sigma2 only}
    H =\frac{3}{4}  \kappa \, \Sigma_{2} \pm \sqrt{\frac{\kappa \rho}{3} + \frac{13 \kappa^2 \Sigma_{2}^2}{16} +\frac{\kappa \, \dot{\Sigma}_{2}}{2} \, } \,.
\end{equation}

\subsubsection{Assuming that shear is proportional to energy density}
\label{subsec: S2 sqrt rho}

Motivated by the dimensional considerations as well as by the calculation in the previous subsection, let us consider again a fixed proportionality relation between the second shear components and energy density,
\begin{equation}
\label{eq: Sigma2 b2}
   \Sigma_2 = b_2 \sqrt{\frac{ \rho}{3 \kappa}} \,.
\end{equation}
By substituting in $\dot{H}$ from the second Friedmann equation \eqref{eq: FR2 Sigma2 only} and then $H$ from the first Friedmann equation \eqref{eq: FR1 Sigma2 only} into the continuity equation \eqref{eq: continuity Sigma2 only}, while replacing the shear contributions by energy density \eqref{eq: Sigma2 b2}, we obtain
\begin{align}
\label{eq: continuity Sigma2 b}
  \ddot{\rho} =& \, \frac{\dot{\rho}^{2}}{2 \rho} - \frac{\sqrt{3 \kappa \rho} \dot{\rho} \left(37 b_{2}^{2} + 16\right)}{12 b_{2}} - \frac{\kappa \rho^{2} \left(13 b_{2}^{2} + 16\right)}{2} 
 \mp \frac{\left(7 \sqrt{3} b_{2} \dot{\rho} + 12 \sqrt{\kappa} (\mathrm{w} + 2 b_{2}^{2} + 1) \rho^{\frac{3}{2}}\right)}{12 \sqrt{\rho} b_{2}} \sqrt{4 \sqrt{3 \kappa \rho} b_{2} \, \dot{\rho} + (13 b_{2}^{2} + 16) \kappa \rho^{2}} \,.
\end{align}
Obviously, the solutions are real if $H$ in \eqref{eq: H Sigma1 only} is real, hence
\begin{align}
\label{eq: Sigma2 b physical}
    \dot{\rho} > - \frac{(16+13b_2^2)\sqrt{\kappa \rho}\rho}{12 b_2} \,.
\end{align}

\begin{figure}[t]
	\centering \includegraphics[width=0.46\textwidth]{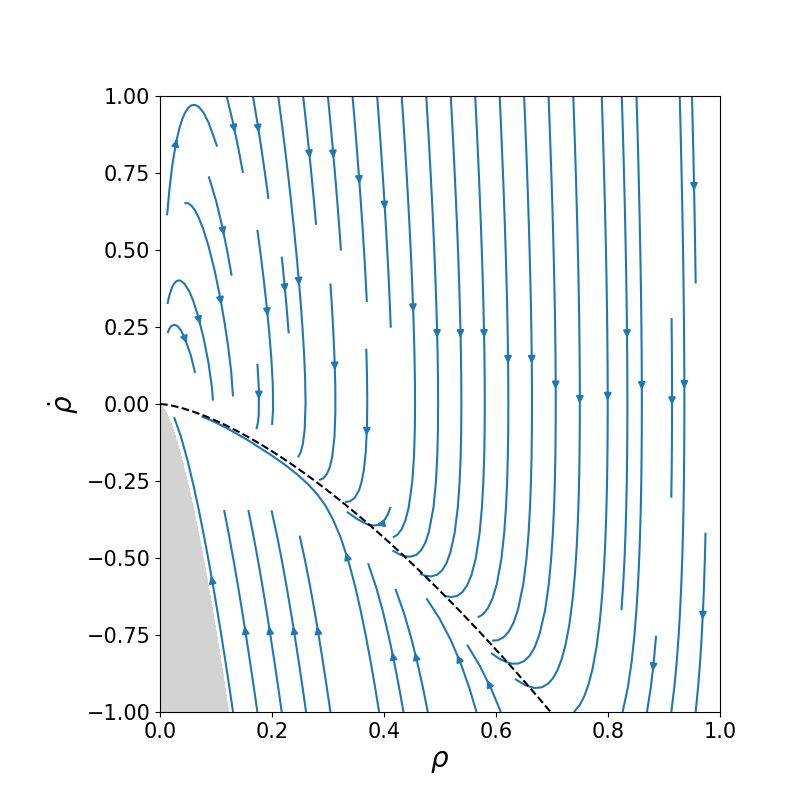}\label{fig: Shear2 b positive}
	\includegraphics[width=0.46\textwidth]{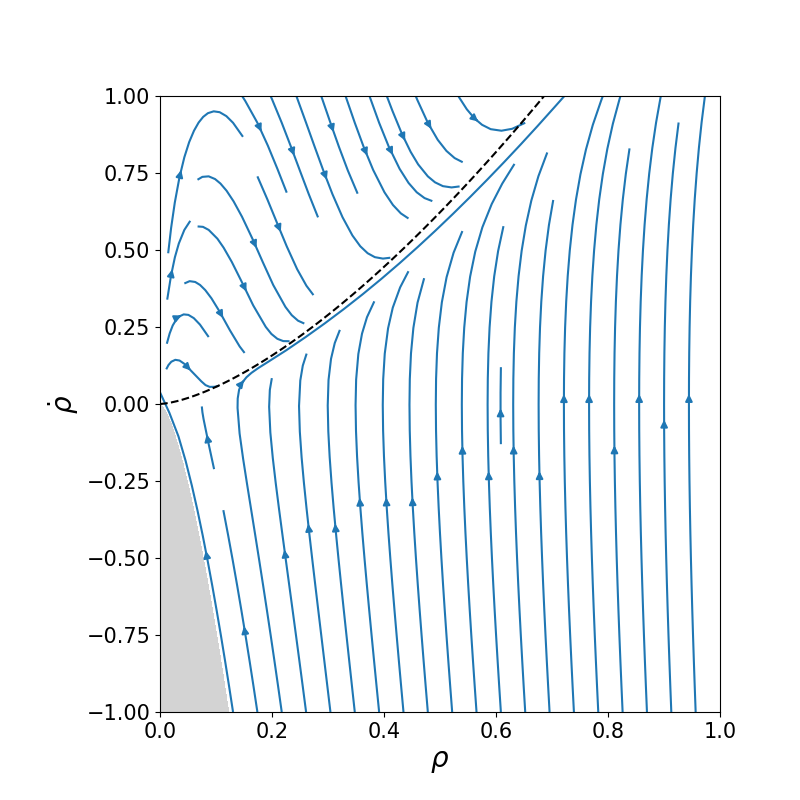}\label{fig: Shear2 b negative}
    \caption{Phase portraits of the  density equation \eqref{eq: continuity Sigma2 b} where the second shear component is proportional to matter density \eqref{eq: Sigma2 b2}. The parameter values are taken to be $b_2=0.1$, $\w=0$, and the left (right) plot corresponds to the positive (negative) sign in the Hubble expression \eqref{eq: FR1 Sigma2 only}. The dashed line corresponds to a ``slow roll'' approximation given by \eqref{eq: density GR eq}, \eqref{eq: Sigma2 b w_rho}, and the grey area marks the unphysical region that violates \eqref{eq: Sigma2 b physical}.}
    \label{fig: Shear2 b}
\end{figure}

Eq.\ \eqref{eq: continuity Sigma2 b} is a complicated second-order differential equation for the density, and can be only solved numerically. Therefore, we have depicted a sample of solutions on a phase diagram, calculated for some parameter values on Fig.\ \ref{fig: Shear2 b}. Here we notice that all trajectories converge to a ``master'' trajectory, rather like in the case of ``slow roll'' dynamics in inflationary cosmology \cite{Liddle:1994dx, Jarv:2021qpp, Jarv:2024krk}. Indeed, inspired by the ``slow roll'' conditions in inflation, in the Eq.\ \eqref{eq: continuity Sigma2 b} we can drop the $\ddot{\rho}$ term and expand for small values of $\dot{\rho}$, which to the first nontrivial order gives precisely the familiar GR type expression \eqref{eq: density GR eq}, albeit the density index is
\begin{align}
\label{eq: Sigma2 b w_rho}
    1+\w_\rho =& \frac{2 \left(13 b_{2}^{2} + 16\right) \left(b_{2} \sqrt{13 b_{2}^{2} + 16} \pm 2 \left(\mathrm{w} + 2 b_{2}^{2} + 1\right)\right)}{ b_{2} \left(24 \mathrm{w} + 139 b_{2}^{2} + 136\right) \pm \sqrt{13 b_{2}^{2} + 16} \left(37 b_{2}^{2} + 16\right)}
\end{align}

It is nice to observe that on the phase diagrams \ref{fig: Shear2 b}, the curve given by \eqref{eq: density GR eq}, \eqref{eq: Sigma2 b w_rho} approximates the full nonlinear ``master trajectory'' very well. This means that for a large part of the history the solutions of Eq.\  \eqref{eq: continuity Sigma2 b} follow an approximation similar to the GR solution \eqref{eq: rho(t) GR} with a modified index \eqref{eq: Sigma2 b w_rho} only. Here, one should note that it is not always guaranteed that this the ``master solution'' and its ``slow roll'' approximation fall into the physical domain \eqref{eq: Sigma2 b physical}. For instance, if the parameter $b_2$ is negative, the physical domain is rather narrow and the ``slow roll'' solutions might reside in the unphysical domain instead  (as can happen in the inflationary models too \cite{Jarv:2021ehj}).

We can substitute the ``slow roll'' approximate $\rho(t)$ solution into the $H$ expression \eqref{H Sigma2 only} and apply the assumption that $\dot{\rho}$ is small. The result is again congruent with the GR counterpart \eqref{eq: H GR)} with different integration constants which absorb the effect of shear,
\begin{subequations}
\label{eq: H0 weff b2}
\begin{align}\label{eq: Sigma2 b H0}
  H_0 &= \left( \pm \frac{ \sqrt{13 b_2^2+16}}{4}+\frac{3 b_2}{4}+\frac{3 b_2 (\w_\rho+1)}{2 \sqrt{13 b_2^2+16}}\right) \sqrt{\frac{\kappa \rho_0}{3}} \,,\\
  \label{eq: Sigma2 b 1+weff}
    1+\weff &= \pm \frac{\sqrt{3 \kappa \rho_0}}{3 H_0} \left( 1+\w_\rho \right) \,.
\end{align}
\end{subequations}
One can also check the latter result from the definition of $\weff$ \eqref{eq: w_eff definition general} using the expressions of $H^2$ \eqref{eq: FR1 Sigma2 only}, $\dot{H}$ \eqref{eq: FR2 Sigma2 only}, and $\dot{\rho}$ in the slow roll approximation \eqref{eq: density GR eq}, \eqref{eq: Sigma2 b w_rho}.

Moreover we can find the expression for $\w_h$ from \eqref{eq: w_h Sigma2 only} using H from \eqref{H Sigma2 only} and $\rho$ and its derivative via the expressions that lead to \eqref{eq: Sigma2 b w_rho}:
\begin{equation}\label{whlast}
    \w_h= \pm \frac{32 +4 \left(b_2^2-8\right) \w + 203 \,b_2^4+232\, b_2^2 \pm \left(56\, b_2^3+32\, b_2\right) \sqrt{13 \,b_2^2+16}}{96-12 \left(b_2^2+8\right) \w + 470 \,b_2^4+560\, b_2^2 \pm \left(131\, b_2^3+80\, b_2\right) \sqrt{13 b_2^2+16}} \,.
\end{equation}

These expressions are rather intractable but make more sense in the $b_2\ll 1$ limit, whereby at the lowest order \eqref{eq: Sigma2 b w_rho}, \eqref{eq: H0 weff b2} reduce to
\begin{subequations}
\label{eq: S2 b small}
\begin{align}
    \w_\rho &\approx \w \mp \frac{(3\w^2 + 20\w +1)b_2}{8} \,,\\
    \weff &\approx \w \mp\frac{(5\w+1)b_2}{2} \,,\\
    H_0 &\approx \left( \pm 1 - \frac{3(\w-1)b_2}{8} \right) \sqrt{\frac{\kappa \rho_0}{3}} \,,\\
    \w_h &\approx \frac{1}{3}-\frac{2 \, b_2}{9 (\w-1)}.
\end{align}
\end{subequations}
Thus, in a reasonable approximation, the effects of a fluid which carries the second shear hypermomentum can also be 
encapsulated by the effective indexes that modify the general relativistic laws of density dilution and space expansion.

\subsubsection{Assuming constant hypermomentum index}
\label{subsec: shear2 both w constant}

We will now assume that $\w$ and $\w_h$ are constant. Again, as done for spin \ref{subsec: spin both w constant}, the constant $\w_h$ hypothesis can be justified looking at the previous section Eq.\ \eqref{whlast}.
In this case it is possible to follow the same procedure as in Sec.\ \ref{subsec: spin both w constant} and \ref{subsubsec: separate conservation laws shear1}. Taking H from \eqref{eq: w_h Sigma2 only} and inserting it into \eqref{eq: continuity Sigma2 only} and \eqref{eq: FR1 Sigma2 only} we obtain 
\begin{align}
\label{dotrhosigma21}
    \dot{\rho} &= -\frac{3 \kappa}{2} \frac{\w_h-1}{3 \w_h-1} \frac{\Sigma_2^2(\rho)}{\frac{\dd\Sigma_2}{\dd\rho } } \pm 
    \sqrt{\frac{3\kappa}{2}}  \sqrt{2 \rho + \frac{3\kappa \Sigma_2^2(\rho)}{3 \w_h-1}} \frac{\Sigma_2(\rho)}{\frac{\dd\Sigma_2}{\dd\rho } } \,, \\
\label{dotrhosigma22}
    \dot{\rho} &=-\frac{3 \kappa}{2} \frac{(\w_h-1) \left(3 \kappa (\w_h+1) \Sigma_2^2(\rho)+2 (\w+1) (3 \w_h-1) \rho\right) \Sigma_2^2(\rho) }{ (3 \w_h-1) \left(3 \kappa (\w_h-1) \Sigma_2^2(\rho) \frac{\dd\Sigma_2}{\dd \rho } +2 (3 \w_h-1) \left(  (\w+1) \rho \frac{\dd\Sigma_2}{\dd\rho }-\Sigma_2(\rho)\right)\right)}
\end{align}
by expecting that the spin is a function of matter density, $\Sigma_2(\rho)$. 
Equating the r.h.s.\ of \eqref{dotrhosigma21} and \eqref{dotrhosigma22} 
we obtain
\begin{align}
\label{sigma2rhoderivat}
   \frac{\dd\Sigma_2}{\dd\rho } &= - \frac{ \left(   (\w_h-1) \mu \Sigma_2 \mp \sqrt{3 \kappa} \sqrt{\mu \Sigma_2^2+\rho } \right) \Sigma_2(\rho)}{2 \mu^2 (\w_h-1) \Sigma_2^3(\rho) 
   \pm \sqrt{3 \kappa}  \left((\w+1) \rho+(\w_h-1) \mu \Sigma_2^2\right) \sqrt{\mu \Sigma_2^2(\rho)+\rho }}
\end{align}
where 
\begin{align}
\mu & =\frac{3 \kappa}{2}\frac{1}{(3 \w_{h}-1)}
\end{align}
and $\w_h\neq 1/3$. 
By using the ansatz $\Sigma_2=\sqrt{\rho} f$ it is possible to write the equation in a separable form like Eq.\ \eqref{separdiff} 
before, and integrate it directly. However, the result is given as a sum of logarithms and linear terms, 
which makes the analytic extraction of $f$ not very illuminating.

If we assume that the shear contribution is much smaller compared to the density, $\frac{\kappa \Sigma_2^2}{\rho} \ll 1$, we can expand the equation above and keep only the leading terms, obtaining
\begin{align}\label{version2}
    \frac{\dd \Sigma_2}{\dd \rho} &\approx \frac{\Sigma_{2}{\left(\rho \right)}}{ \left(\w + 1\right) \rho} \mp  \frac{\sqrt{3 \kappa} \left(\w_h - 1\right) \Sigma_{2}^{2}{\left(\rho \right)}}{2 \left(\w + 1\right) \left(3 \w_h - 1\right) \rho^{\frac{3}{2}}}
\end{align}
This is structurally analogous to Eq.\ \eqref{sigma1approx} and can be readily integrated to obtain
\begin{align}\label{version2sol}
    \Sigma_2(\rho) &=\mp \frac{ \frac{(3 \w_h -1)(\w-1)}{\w_h-1} }{1-\left(1 \pm \frac{(3 \w_h -1)(\w-1)}{\w_h-1} \sqrt{\frac{\rho_0}{3\kappa (\Sigma_{2,0})^2}} \right)\left(\frac{\rho}{\rho_0} \right)^{\frac{\w-1}{2(\w+1)}}} \sqrt{\frac{\rho}{3\kappa}} \,,
\end{align}
where the integration constant is fixed so that at some arbitrary time $\Sigma_{2,0}=\Sigma_2(\rho_0)$. For late universe with small $\rho$, the regime of $\kappa \Sigma^2_2 \ll \rho$ can be maintained if $-1 < \w < 1$, i.e.\ for practically all realistic barotropic fluids.

\subsubsection{Assuming separate conservation laws for energy and second shear component}

Like in Sec.\ \ref{subsec: separate conservation spin} and \ref{subsubsec: separate conservation laws shear1}
we can assume that both sides of \eqref{eq: continuity Sigma2 only} hold separately, that is
\begin{equation}\label{241}
   \dot{\rho} + 3 H \rho \, (1+\w)=0
\end{equation}
and
\begin{equation}\label{242}
  3 \, \Sigma_2 \dot{H} +\frac{\dd \Sigma_2}{\dd \rho} \dot{\rho} \, (7 H+ \kappa  \Sigma_2)+3 \kappa H \Sigma_2^2+12 H^2 \Sigma_2+\frac{\dd \Sigma_2}{\dd \rho} \ddot{\rho} +\frac{\dd^2 \Sigma_2}{\dd \rho^2} \dot{\rho}^2=0 \,,
\end{equation}
where $\Sigma_2$ was taken to be a function of $\rho$.
We can use \eqref{eq: FR2 Sigma2 only} and \eqref{H Sigma2 only} to eliminate $\dot{H}$ and $H$ in \eqref{241} and \eqref{242}. The latter is a second-order equation, but as before, we can invoke the ``slow roll'' approximation setting $\ddot{\rho} \approx 0$ and keeping only linear terms in the expansion of $\dot{\rho}\ll1$. To be able to solve the combined equation of \eqref{241} and \eqref{242} with these substitutions, we also need to take $\kappa \Sigma_2^2 \ll \rho$ as in similar cases before. 
The resulting equation
\begin{align}
    \frac{\dd \Sigma_2}{\dd \rho} &\approx - \frac{(3\w-5)\Sigma_2}{14(\w+1)\rho} \pm \frac{3\sqrt{3 \kappa}(9\w^2+86\w-19)(\Sigma_2)^2}{784(\w+1)\rho^{3/2}}
\end{align}
is structurally the same as \eqref{sigma1approx} and \eqref{version2}. Thus it is easy to write out the solution
\begin{equation}
    \Sigma_2 \approx \pm     
    \frac{\frac{112 (5 \w+1)}{3 \left(9 \w^2+86 \w-19\right)}}{ 1 + \left( - 1 \pm \frac{112 (5 \w+1)}{3 \left(9 \w^2+86 \w-19\right)} \frac{\sqrt{\rho_0}}{\sqrt{3\kappa} \Sigma_{2,0}}\right)\left(\frac{\rho }{\rho_0}\right)^{\frac{5 \w+1}{7 (\w+1)}}}    
    \sqrt{\frac{\rho }{3 \kappa}} 
\end{equation}
For late universe, when $\rho$ is small, the condition $\kappa \Sigma_2^2 \ll \rho$ is mantained if $-1<\w<-\frac{1}{5}$. In the other cases of $\w< -1$ or $\w>-\frac{1}{5}$ the shear enters in a regime $\kappa \Sigma_2^2 \sim \rho$ that has been studied in section \ref{subsec: S2 sqrt rho}.

\subsubsection{Assuming de Sitter expansion}\label{desittershear2}

Finally, let us assume de Sitter expansion, i.e., $H=H_*$.
By subtracting equation \eqref{eq: FR1 Sigma2 only} from equation \eqref{eq: FR2 Sigma2 only}, we get the relation
\begin{equation}
\label{Sigma_2:deSitter:1}
    (1+\w ) \rho + 6 H_{*} \Sigma_{2} + \frac{3}{2} \kappa \Sigma_{2}^{2} +2 \dot{\Sigma}_{2} = 0 \,. 
\end{equation}
Differentiating this expression and using the result in the continuity equation 
\eqref{eq: continuity Sigma2 only}, we get an evolution equation for $\Sigma_{2}(t)$,
\begin{equation}
\label{Sigma_2_deSitter:1}
    \ddot{\Sigma}_{2} + 3 \, H_{*} \, \dot{\Sigma}_{2} +  \frac{3(\w-1)}{(3\w-1)} \kappa \, \Sigma_{2} \, \dot{\Sigma}_{2} =0\, .
\end{equation}
We can see that Eq.\ \eqref{Sigma_2_deSitter:1} is either trivially satisfied if $\Sigma_2$ is constant, $\Sigma_2 = C_0$, or for $\Sigma_{2}\neq 0$ the first integration of the above yields
\begin{equation}\label{firstint}
   \dot{\Sigma}_{2} + 3  H_{*} \Sigma_{2} +\frac{\alpha  \kappa }{2}\Sigma_{2}^2 = c_1 \,,
\end{equation}
where again we denote $\alpha=  \frac{3(\w-1)}{(3\w-1)}$. Separating the variables and rearranging the previous equation gives
\begin{equation}
    \frac{d  \Sigma_{2}}{\left(\frac{3 H_{*} }{\alpha  \kappa}+\Sigma_{2}\right)^2 -  \Gamma}=- \frac{\alpha  \kappa}{2} \text{dt}
\end{equation}
where the solution depends on the value of the constant 
\begin{equation}
    \Gamma = \frac{2}{\alpha  \kappa} \left(c_1 +\frac{9 H_{*}^2}{2 \alpha  \kappa}\right) \,.
\end{equation}
We can distinguish the following cases.

If $\Gamma>0$, and denoting $\Gamma=\gamma^{2}$, we obtain the solution
\begin{subequations}
\begin{equation}\label{tanhS}
   \Sigma_{2}= -\frac{3 H_{*}}{\alpha  \kappa} + \gamma \, \tanh\left(\frac{\alpha  \kappa }{2} \gamma t+c_{2}\right) 
\end{equation}
with the constants
\begin{align}
    c_1 &=\dot{\Sigma}_{2,0} + 3  H_{*} \Sigma_{2,0}+\frac{\alpha  \kappa}{2} \Sigma_{2,0}^2 \,, \\
    c_2 &= \tanh^{-1} \left(\frac{1}{\gamma} \left(\frac{3 H_{*}}{\alpha \kappa}+\Sigma_{2,0}\right)\right) -\frac{\alpha  \kappa}{2} t_0 \gamma \,. \label{c2}
\end{align}
\end{subequations}
For $\Gamma<0$ we simply set $\Gamma = (i \gamma)^2$ (and $\gamma$ real) to obtain instead the solution
\begin{subequations}
\begin{equation}\label{ds22}
   \Sigma_{2}= -\frac{3 H_{*}}{\alpha  \kappa} - \gamma \, \tan\left(\frac{\alpha  \kappa }{2} \gamma t+c_{2}\right)
\end{equation}
where
\begin{equation}
    c_2 = \tan^{-1}\left(\frac{1}{\gamma} \left(\frac{3 H_{*}}{\alpha  \kappa} -\Sigma_{2,0}\right)\right) -\frac{\alpha  \kappa}{2} t_0 \gamma \,.
\end{equation}
\end{subequations}
This branch of the solutions is clearly unphysical since shear blows up in finite times.
\newline Finally, for $\Gamma=0$ the solution is
\begin{subequations}\label{ExpSS}
\begin{align}
  \Sigma_{2} &=  -\frac{3 H_{*}}{\alpha  \kappa} +\frac{1}{\frac{\alpha \kappa}{2} t+c_{2}}  \,
\end{align}
with
\begin{align}
  c_2 &= \frac{1}{2} \alpha   \kappa \left(-t_0+\frac{2}{3 H_{*} +\alpha   \kappa \Sigma_{2,0}}\right) \,.
\end{align}
\end{subequations}
In conclusion, we see that a de Sitter universe is supported by a non-constant shear hypermomentum $\Sigma_{2}$ which is given by either (\ref{tanhS}) or (\ref{ExpSS}) depending on the parameters. For both of these solutions, at late times, shear approaches a constant value.

\section{Multiple component hyperfluid}
\label{sec: 6 Multiple component hyperfluid}

Finding general analytic solutions for the system \eqref{eq: FLRW equations general} with all hyperfluid components present is practically impossible. However, in the previous section, we developed several approaches and typical simplification assumptions that allowed us to solve the system. In this section we exemplify how these methods are useful in the multiple component case as well.

\subsection{All hyperfluid components proportional to energy density}

In the first example, let us take all hypermomentum quantities to be proportional to the square root of the energy density with different parameters,
\begin{align}\label{sigmashear1and2}
    \sigma &= b \sqrt{\frac{3 \rho}{\kappa}} \,, \qquad \Sigma_1 = b_1 \sqrt{\frac{\rho}{3 \kappa}} \,, \qquad \Sigma_2 = b_2 \sqrt{\frac{\rho}{3 \kappa}} \,.
\end{align}
Then the continuity equation \eqref{eq: continuity eq} with H substituted in from \eqref{eq: FR1} reads as
\begin{align}
\label{eq: ddotrho b, b1, b2}
    \ddot{\rho} &= - \frac{\beta_{1} \beta_{3} \kappa \rho^{2}}{12 b_{2}} \mp \frac{\beta_{2} \sqrt{\kappa} \rho \sqrt{\beta_{1} \kappa \rho^{2} + 4  b_{2} \sqrt{3 \kappa \rho} \dot{\rho} }}{12 b_{2}} + \frac{\dot{\rho}^{2}}{2 \rho} - \frac{\sqrt{3} \dot{\rho} \left( (\beta_{1} + 4 \beta_{3} b_{2} ) \sqrt{\kappa} \rho \pm (2 b_{1} +7 b_{2} ) \sqrt{\beta_{1} \kappa \rho^{2} + 4  b_{2} \sqrt{3 \kappa \rho} \dot{\rho} } \right)}{12 \sqrt{\rho} b_{2}} \,,
\end{align}
where the parameters have been grouped into the combinations
\begin{subequations}
\begin{align}
    \beta_1 &=12 b b_{2} + 4 b_{1}^{2} + 12 b_{1} b_{2} + 13 b_{2}^{2} + 16 \,, \\
    \beta_2 &= 12 \w + 6 b b_{1} + 21 b b_{2} + 8 b_{1}^{2} + 28 b_{1} b_{2} + 24 b_{2}^{2} + 12 \,, \\
    \beta_3 &= 3 b + 4 b_{1} + 6 b_{2} \,.
\end{align}
\end{subequations}
The solutions are real if $H$ is real, hence
\begin{align}
    \dot{\rho} > - \frac{\beta_1 \sqrt{3 \kappa \rho} \rho}{12 b_2} \,.
\end{align}
We can notice that Eq.\ \eqref{eq: ddotrho b, b1, b2} has the same structure as Eq.\ \eqref{eq: continuity Sigma2 b} and coincides with it when $b=b_1=0$. Therefore the same ``slow roll'' approximation works, whereby we can drop the $\ddot{\rho}$ term and expand for small values of $\dot{\rho}$, which to the first nontrivial order gives precisely the familiar GR type expression \eqref{eq: density GR eq}, whereby the density index is 
\begin{align}
\label{eq: 1+w_rho b, b_1, b_2}
    1+\w_\rho &= \frac{\beta_{1}^{\frac{3}{2}} \beta_{3} \pm \beta_{1} \beta_{2}}{3 \left(2 \beta_{1} b_{1} + 7 \beta_{1} b_{2} + 2 \beta_{2} b_{2} \pm (\beta_{1} + 4 \beta_{3} b_{2}) \sqrt{\beta_{1}}\right)} \,.
\end{align}
Like in Sec.\ \ref{subsec: S2 sqrt rho}, in this approximation, the expansion law also follows the GR rule \eqref{eq: H GR)} with
\begin{equation}
\label{eq: H_0 b, b_1, b_2}
    H_0=\left(6 b + 2 b_1 + 3 b_2 \pm \sqrt{\beta_1} -\frac{6 b_2 (\w_\rho+1)}{\sqrt{\beta_1}}\right) \frac{1}{4} \sqrt{\frac{\kappa \rho_0}{3}} 
\end{equation}
In the limit where all the parameters $b, b_1, b_2$ can be all considered to be small, we obtain the sum of earlier results \eqref{eq: w_rho, w_eff, H_0 small b}, \eqref{eq: S1 b1 small}, \eqref{eq: S2 b small}, 
\begin{subequations}
\label{eq: w_rho, w_eff, H_0 in b, b_1, b2}
\begin{align}
   \w_\rho &\approx \w \pm b \mp \frac{(3\w-5)b_1}{6} \mp \frac{(3\w^2 + 20\w +1)b_2}{8} \,,\\
   \weff &\approx \w \mp \frac{(3 \w +1)b}{2} \mp \frac{(3\w-1) b_1}{3} \mp\frac{(5\w+1)b_2}{2} \,,\\
   H_0 &\approx \left( \pm 1 + \frac{3b}{2} +\frac{b_1}{2} -  \frac{3(\w-1)b_2}{8} \right) \sqrt{\frac{\kappa \rho_0}{3}} \,.
\end{align}
\end{subequations}
It is interesting that for certain combinations of model parameters we can exactly reproduce the general relativistic evolution. For instance, taking
\begin{align}
    b &= \frac{(\w^2+2\w+1)b_2}{2\w-2} \,, \qquad b_1 = -\frac{(3\w^2+18\w+3)b_2}{4\w-4} 
\end{align}
in \eqref{eq: w_rho, w_eff, H_0 in b, b_1, b2} would make the first order corrections to vanish and give $\w_\rho \approx \w$, $\weff \approx \w$, $H_0 \approx \sqrt{\frac{\kappa \rho_0}{3}}$. In the case of dust matter with $\w=0$ a combination like $b=-0.1$, $b_1=0.15$, $b_2=0.2$ would do it. Similarly, the full solutions with \eqref{eq: 1+w_rho b, b_1, b_2}, \eqref{eq: H_0 b, b_1, b_2}, \eqref{eq: Sigma2 b 1+weff} can also reduce to exactly general relativistic evolution, although the relationships between the parameters would now involve several square roots. In any case at least one of the parameters $b$, $b_1$, $b_2$ should be of the opposite sign compared to the others.

\subsection{Assuming all hyperfluid components proportional to energy density and  constant hypermomentum index}

In the previous section, we were able to solve the system in the slow roll and small $b$'s approximation. An alternative option is to still take \eqref{sigmashear1and2} but instead assume that the hypermomentum index $\w_{h}$ is constant.
The hypermomentum effective density and pressure \eqref{rhohph} with \eqref{sigmashear1and2} are
\begin{align}
\label{rhoh3}
  \rho_h &= \frac{3 \sqrt{3}}{2 \sqrt{\kappa}} \left(2 b +\frac{2 b_1}{3}+b_2\right) H \sqrt{\rho} + \frac{3}{4} \left(-3 b^2-2 b (b_1+b_2)+\frac{b_2^2}{3}\right) \rho  +\frac{\sqrt{3} \, b_2}{4 \sqrt{\kappa}} \frac{\dot{\rho}}{\sqrt{\rho}} \, , \\
\label{ph3}
   p_h &= \frac{\sqrt{3}}{2 \sqrt{\kappa}} \left(-4 b +\frac{2 b_1}{3}+b_2\right) H \sqrt{\rho} +
    \frac{1}{4} \left(3 b^2-2 b (b_1+b_2)+\frac{4 b_1 b_2}{3}+b_2^2\right) \rho + 
    \frac{\sqrt{3}}{4 \sqrt{\kappa}} \left(\frac{b_2}{3}-2 b \right)\frac{\dot{\rho}}{\sqrt{\rho}} \,.
\end{align}
For constant $\w_h$ in \eqref{eq: w_h definition} these imply
\begin{equation}\label{eqgamma}
   \gamma_1  \frac{\dot{\rho}}{\sqrt{\rho}}+ \gamma_2  \sqrt{\kappa} \rho + \gamma_3 H \sqrt{\rho}=0
\end{equation}
where the coefficients are
\begin{subequations}
\begin{align}
   \gamma_1 &=2 b -\frac{b_2}{3}+b_2 \w_h \,, 
\\
 \gamma_2 &=\sqrt{3} \left( - b^{2} + \frac{2 b b_{1}}{3} + \frac{2 b b_{2}}{3} - \frac{4 b_{1} b_{2}}{9} - \frac{b_{2}^{2}}{3} - \mathrm{w}_h \left(3 b^{2} + 2 b b_{1} + 2 b b_{2} - \frac{b_{2}^{2}}{3}\right) \right)\,,
\\
   \gamma_3&= 8 b - \frac{4 b_{1}}{3} - 2 b_{2} + \mathrm{w}_h \left(12 b + 4 b_{1} + 6 b_{2}\right)\,.
\end{align}
\end{subequations}
Substituting the ratio $\frac{\dot{\rho}}{\sqrt{\rho}}$ from \eqref{eqgamma} into \eqref{rhoh3} we obtain
\beq\label{rhoh4}
\rho_{h}=\gamma_{4} \, \rho+\gamma_{5} \sqrt{\frac{3 \rho}{\kappa}} H
\eeq
where
\begin{subequations}
\begin{align}
    \gamma_{4} &= - \frac{9 b^{2}}{4} - \frac{3 b b_{1}}{2} - \frac{3 b b_{2}}{2} + \frac{b_{2}^{2}}{4} - \frac{\sqrt{3} \gamma_{2} b_{2}}{4 \gamma_{1}} \,,
\\
    \gamma_{5} &= 3 b + b_{1} + \frac{3 b_{2}}{2} - \frac{\gamma_{3} b_{2}}{4 \gamma_{1}}\,.
\end{align}
\end{subequations}
A further substitution of \eqref{rhoh4} into the first Friedmann equation \eqref{eq: FR1} brings it to the form
 \begin{equation}
     H^2= \frac{\kappa}{3} (1 + \gamma_4) \rho +  \gamma_5 \sqrt{\frac{\kappa \rho}{3}} H \,.
 \end{equation}
This is a quadratic equation in $H$ with solutions
\beq\label{Hpm}
H=\lambda_{\pm} \sqrt{\frac{\kappa \rho}{3}}
\eeq
where
 \begin{equation}
    \lambda_{\pm}= \frac{1 }{2} \left( \gamma_5 \pm \sqrt{4 \gamma_4+ \gamma_5^2+4}\right) \,.
 \end{equation}
Combining this solution with \eqref{eqgamma} we obtain the equation
 \begin{equation}
\dot{\rho} +\frac{1}{\gamma_1} \left(\gamma_2 \sqrt{\kappa} +\gamma_3  \sqrt{\frac{\kappa}{3}} \lambda_{\pm} \right) \rho^{3/2}=0
 \end{equation}
which is in the same form as \eqref{eq: density GR eq} in general relativity, only the density index is given by
\begin{equation}\label{wrholast}
    \w_\rho + 1 = \pm \frac{1}{\gamma_1 \sqrt{3}} (\gamma_2+ \gamma_3 \frac{1}{\sqrt{3}} \lambda_{\pm}) \,.
\end{equation}
Substituting the ratio $\frac{\dot{\rho}}{\sqrt{\rho}}$ from \eqref{eqgamma} into \eqref{ph3} we obtain
\beq\label{ph4}
p_{h}=\gamma_{6} \, \rho+ \gamma_{7} H \sqrt{\frac{3 \rho}{\kappa}}
\eeq
where
\begin{subequations}
\begin{align}
    \gamma_{6} &= \frac{3 b^{2}}{4} - \frac{b b_{1}}{2} - \frac{b b_{2}}{2} + \frac{b_{1} b_{2}}{3} + \frac{b_{2}^{2}}{4} + \frac{\sqrt{3} \gamma_{2} b}{2 \gamma_{1}} - \frac{\sqrt{3} \gamma_{2} b_{2}}{12 \gamma_{1}} \,,
\\
    \gamma_{7} &= - 2 b + \frac{b_{1}}{3} + \frac{b_{2}}{2} + \frac{\gamma_{3} b}{2 \gamma_{1}} - \frac{\gamma_{3} b_{2}}{12 \gamma_{1}} \,.
\end{align}
\end{subequations}
A further substitution of \eqref{ph4} into the second Friedmann equation \eqref{eq: FR2 GR} gives
\begin{equation}\label{32r}
2 \dot{H}+3 H^2 + \kappa \w \rho + \kappa \left(  \gamma_7 H \sqrt{\frac{3 \rho}{\kappa}}+ \gamma_6 \rho \right)=0
\end{equation}
where we have the expression relating H and $\rho$ in \eqref{Hpm}. Thus it is possible to compare the \eqref{32r} 
with \eqref{wefforigin} to identify
\begin{align}
    \weff &= \frac{\w + \gamma_6 + \lambda_\pm \gamma_7}{\lambda_\pm^2} \,.    
\end{align}
Let us note that we have a different result with respect to the previous section because we are assuming $\w_h$ to be constant, 
and this strong assumption does not leave any freedom to proceed, for example, considering an expansion for small b's.

\subsection{Equal shear components, no spin,  constant indices}

In the third example, we take $\sigma=0$ and assume $\Sigma_1=\Sigma_2=\Sigma$.
The equations \eqref{eq: FR1}, \eqref{eq: FR2}, \eqref{eq: continuity eq},
and \eqref{eq: w_h definition} become:
\begin{align}
\label{firstfr23}
    12 H^2 &=\kappa \left(30 H \Sigma+3 \kappa \Sigma^2+4 \rho+6 \dot{\Sigma}\right) \,, \\
    2 \dot{H} +3 H^2 &=-\frac{1}{4} \kappa \left(10 H \Sigma+7 \kappa \Sigma^2+2 \dot{\Sigma}+4 \rho \, \w\right) \,, \\
\label{continuity23}
  \dot{\rho} + 3 H \rho (\w+1) &=-\frac{3}{2} \left(\Sigma \left(5 \dot{H} + \kappa \dot{\Sigma}\right)+ H \left(5 \kappa \Sigma^2+9 \dot{\Sigma}\right)+20 H^2 \Sigma+\ddot{\Sigma}\right) \,,
\end{align}
and
\begin{equation}\label{wh23}
   \w_h = \frac{10 H \Sigma + 7 \kappa \Sigma^2+2 \dot{\Sigma}}{30 H \Sigma+3 \kappa \Sigma^2+6 \dot{\Sigma}} \,,
\end{equation}
while the expression for H can be found solving \eqref{firstfr23},
\begin{equation}
\label{eq: H Sigma}
    H = \frac{5}{4} \kappa \Sigma \pm \sqrt{\frac{29}{16} \kappa^2 \Sigma^2+\frac{1}{3} \kappa \rho+\frac{1}{2} \kappa \dot{\Sigma}} \,.
\end{equation}

If we assume that the hyperfluid index \eqref{wh23} is constant along with the matter index $\w$, we can proceed like in Sec.\ \ref{subsec: shear2 both w constant}. On the one hand, it is possible to
express $H$ from \eqref{wh23} and equate it with the $H$ expression from the first Friedmann equation \eqref{eq: H Sigma}. Considering that $\Sigma$ is a function of $\rho$ (thus $\dot{\Sigma}=\frac{\dd \Sigma}{\dd \rho} \dot{\rho}$) this procedure gives
\begin{equation}
    \dot{\rho}= -\frac{\kappa (3 \w_h-7) \Sigma^2}{2 (3 \w_h-1) \frac{\dd \Sigma}{\dd \rho}} \mp \frac{5 \sqrt{\kappa} \Sigma}{ \frac{\dd \Sigma}{\dd \rho}} 
    \sqrt{\frac{9 \kappa \Sigma^2+(6 \w_h-2) \rho}{18 \w_h-6}} \,.
\end{equation}
This is to be compared to the $\dot{\rho}$ we can extract from the continuity equation \eqref{continuity23} where $H$ has been eliminated by using \eqref{wh23},
\begin{equation}
  \dot{\rho}= -\frac{3 \kappa (3 \w_h-7) \Sigma^2 \left(9 \kappa (\w_h+1) \Sigma^2+2 (\w+1) (3 \w_h-1) \rho\right)}{2 (3 \w_h-1) \left(9 \kappa (3 \w_h-7) \Sigma^2 \frac{\dd \Sigma}{\dd \rho} + 6 (\w+1) (3 \w_h-1) \rho \frac{\dd \Sigma}{\dd \rho}+(10-30 \w_h) \Sigma\right)} \,.
\end{equation}
Setting these expressions equal, we obtain
\begin{equation}
\label{eq: Sigma' full}
  \frac{\dd \Sigma}{\dd \rho} =  \frac{ \pm \mu  (3 \w_h -7) \Sigma^2 + 15  \sqrt{3 \kappa } \Sigma \sqrt{\mu \,\Sigma^2+\rho }}{\mp 2 \mu^2 (3 \w_h-7) \Sigma^3 + 3 \sqrt{3 \kappa } \sqrt{\mu \,\Sigma^2 + \rho } \left( \mu (3 \w_h-7) \Sigma^2+3 \rho  (\w+1)\right)}
\end{equation}
where
\begin{equation}
   \mu =  \frac{9 \, \kappa}{2 (3 \w_h -1)} \,.
\end{equation}
In principle, this equation can be integrated by a change of variables like in Sec.\ \ref{subsec: shear2 both w constant}, but getting the explicit form of $\Sigma(\rho)$ from that is difficult.
Instead, we can simplify the equation by taking the limit $\frac{k \Sigma^2}{\rho} \ll 1$ which approximates \eqref{eq: Sigma' full} as
\begin{equation}
   \frac{\dd \Sigma_2}{\dd \rho} \approx \frac{5 \, \Sigma}{3 \rho \,(\w+1)} \mp  \frac{\sqrt{3 \kappa} \, (3 \w_h-7) \,\Sigma^2}{6 \rho ^{3/2} (\w+1)\, (3 \w_h-1)} \,.
\end{equation}
The resulting equation is  structurally similar to the earlier equations for $\Sigma_1$ and $\Sigma_2$, \eqref{sigma1approx} and \eqref{version2}. 
The solution after fixing the integration constant is
\begin{equation}
\Sigma=  \mp \frac{\frac{(3 \w-7) (3 \w_h-1)}{(3 \w_h-7)}}{1 + \left(-1 \mp \frac{\sqrt{\rho_0}}{\sqrt{3 \kappa} \, \Sigma_{2,0}}\frac{ (3 \w-7) (3 \w_h-1)}{(3 \w_h-7) }\right) \left(\frac{\rho }{\rho_0}\right)^{\frac{3 \w-7}{2 (3 (\w+1))}} } \sqrt{\frac{\rho}{3 \kappa}} \,.
\end{equation}
The assumption $\kappa\Sigma^2 \ll \rho$ holds if $-1 < \w < 7/3$, i.e.\ for all realistic barotropic fluids.

\subsection{Constant hypermomentum variables}

Let us now consider the case where all hypermomentum variables are constants for all times, 
namely $\Sigma_{1}=\Sigma_{1}^{0}, \; \Sigma_{2}=\Sigma_{2}^{0}$ and $\sigma=\sigma_{0}$ $\forall t$. 
In this case, the defining equations \eqref{rhohph}, (\ref{pph}) of $\rho_{h}$ and $p_{h}$ take the form
\beq
\rho_{h}=A_{0}+B_{0}H
\eeq
and
\beq
p_{h}=C_{0}+D_{0}H
\eeq
respectively, where 
\begin{subequations}
\begin{align}
    A_{0}&=\kappa \left(- \frac{3 \Sigma_{1,0} \sigma_{0}}{2} + \frac{3 \Sigma_{2,0}^{2}}{4} - 
    \frac{3 \Sigma_{2,0} \sigma_{0}}{2} - \frac{3 \sigma^{2}_{0}}{4}\right) \,, \\  
    B_{0} &=\left(3 \Sigma_{1,0} + \frac{9 \Sigma_{2,0}}{2} + 3 \sigma_{0}\right)\,, \\ 
    C_{0} &= \kappa \left(\Sigma_{1,0} \Sigma_{2,0} - \frac{\Sigma_{1,0} \sigma_{0}}{2} + 
    \frac{3 \Sigma_{2,0}^{2}}{4} - \frac{\Sigma_{2,0} \sigma_{0}}{2} + \frac{\sigma_{0}^{2}}{4}\right) \,, \\
    D_{0} &= \left(\Sigma_{1,0} + \frac{3 \Sigma_{2,0}}{2} - 2 \sigma_{0}  \right)\,.
\end{align}
\end{subequations} 
Combining the above two equations for hypermomentum -- induced density and pressure, we can express
\beq
    p_{h}=\w_{0}\rho_{h}+p_{0}
\eeq
where 
\beq
    \w_{0}=\frac{D_{0}}{B_{0}} \,, \qquad  p_{0}=\frac{B_{0}C_{0}-A_{0}D_{0}}{B_{0}} \,.
\eeq
Also, from the first Friedmann equation \eqref{eq: FR1} and for the given form of $\rho_{h}$ we get
\beq
    \rho= \frac{3}{\kappa}H^{2}-B_{0}H-A_{0} \,.
\eeq
Using the latter, along with the above expressions for the hypermomentum pressure and density, the generalized 
continuity equation (\ref{eq: continuity eq}) becomes
\begin{equation}
   \dot{H} + \frac{1}{2} \kappa (D_{0}-B_{0} \w) H +\frac{3}{2} (\w+1) H^2=\frac{1}{2} \kappa (A_{0} \w- C_{0}) \,.
\end{equation}
Quite interestingly, the form of this equation is identical to \eqref{firstint}, which we encountered in 
sec. \ref{desittershear2} where we studied de-Sitter solutions. 
In fact, there is a duality between $\Sigma_{2}$
appearing there and $H$ here, along with the coefficients of the differential equations. 
The solutions for $H$ then are identical to the solutions \eqref{tanhS}, \eqref{ds22} and \eqref{ExpSS} with the 
mere replacement 
\begin{subequations}
\begin{align}
    \Sigma_{2} \quad  &\leftrightarrow \quad  H \,, \\
    3 H_{*} \quad  &\leftrightarrow \quad  \frac{\kappa}{2}\Big( D_{0}- B_{0} \w \Big) \,, \\
    \alpha \kappa \quad  &\leftrightarrow \quad  3(1+\w) \,, \\
    c_{1} \quad  &\leftrightarrow \quad \frac{\kappa}{2} (- C_{0} + A_{0} \w) \,.
\label{corres}
\end{align}
\end{subequations}
Given these solutions for $H$, it is then trivial to find the expression of the scale factor $a(t)$. 
Setting 
\beq
    \xi_{0}=-\frac{\kappa(D_{0}-\w B_{0})}{6(1+ \w)}\ , \qquad 
    \Gamma=\frac{2 c_{1}}{3(1+\w)}+\xi_{0}^{2}
\eeq
we have the following three cases depending on the value of $\Gamma$. \\

If $\Gamma=\gamma^{2}>0$ we have
\begin{equation}
\label{neg1}
    H =\frac{2 \gamma_0 }{3(\w+1)} \tanh (c_2+\gamma_0 t) + \xi_{0} \ , \qquad  
    \gamma_{0} =\frac{3(1+\w)}{2}\gamma  
\end{equation}
where
\begin{equation}
    c_2 = -\gamma_{0} t_0 + \tanh^{-1}\left(\frac{3 (\w+1) (\xi_{0} - H_0)}{2 \gamma_{0}}\right) \,.
\end{equation}
And solving \eqref{neg1} for scale factor
\begin{equation}
    a(t)= C \Big[ \cosh{(\gamma_{0}t+c_{2})}\Big]^{\frac{2}{3(1+\w)}}e^{\xi_{0}t}
\end{equation}
where $C$ is an integration constant and $c_{2}$ is given by (\ref{c2}) with the correspondences (\ref{corres}). 
We see that, in this case, we have essentially an accelerated expansion. \\
    
If $\Gamma=0$ we have
\begin{equation}
\label{neg2}
    H=\frac{1}{c_{2}+\frac{3}{2} (\w+1) \, t }+\xi_{0}
\end{equation}
where
\begin{equation}
    c_2 =\frac{1}{H_0-\xi_{0} }-\frac{3}{2} (\w+1) \, t_0 \,.
\end{equation}
And solving \eqref{neg2}
\begin{equation}
    a(t)=C\left( \frac{3(1+\w)}{2}t+c_{2}\right)^{\frac{2}{3(1+\w)}}e^{\xi_{0}t} \,,
\end{equation} 
which is again mostly driven by the exponential.\\

If $\Gamma=- \gamma^{2}<0$ we get:
\begin{equation}
\label{neg3}
    H=\frac{2 \gamma_0 }{3 (\w+1)} \tan (-\gamma_0 t-c_2) +\xi_{0}  \,, \qquad  
    \gamma_{0}=\frac{3(1+\w)}{2}\gamma 
\end{equation}
where
\begin{equation}
    c_2 = -\gamma_0 t_0-\tan ^{-1}\left(\frac{3 (\w+1) (H_0+\xi_{0} )}{2 \gamma_0}\right) \,.
\end{equation}
And solving \eqref{neg3}
\beq
    a(t)=C \Big[ \cos{(\gamma_{0} t+c_{2})}\Big]^{\frac{2}{3(1+\w)}}e^{\xi_{0} t} \,.
\eeq
Here we see that we have a bouncing Universe solution.

\subsection{Restoring the usual continuity equation}

If we assume constant $\w_h$ and a dependence $\rho_{h}=\rho_{h}(\rho)$ the generalized continuity equation 
\eqref{eq: continuity eq} takes the form
\beq
\dot{\rho}\Big( 1+\frac{d \rho_{h}}{d \rho} \Big)=
-3H(1 + \w)\rho \Big(1 + \frac{1+\w_{h}}{1+\w}\frac{\rho_{h}}{\rho} \Big) \,.
\eeq 
Requiring that ordinary matter satisfies the usual continuity equation, and we may equate the two big parentheses 
in the above equation to obtain
\beq
\frac{d\rho_{h}}{d \rho}=\frac{1+\w_{h}}{1+\w}\frac{\rho_{h}}{\rho}
\eeq
which can, for constants, barotropic indexes immediately integrate to
\begin{equation}
    \rho_h = \rho_{h, 0}  \, \left( \frac{\rho}{\rho_0}\right)^{\frac{1 + \w_{h}}{1 + \w}} 
    \,= \rho_{h, 0}  \, \left( \frac{a}{a_0}\right)^{-3(1 + \w_h)} 
\label{rrt}
\end{equation}
where $\rho_{h,0}=\rho_{h}(\rho_{0})$ and $\rho_{0}$ being some fixed reference value of the energy density.

Applying the found relationships to the Friedmann equations \eqref{eq: FR1}, we get the expression
\begin{subequations}
\begin{align}
\label{eq:Hlast}
    H^2 & = \frac{\kappa}{3} \left[\rho_0 \left( \frac{a}{a_{0}} \right)^{-3(1+\w)} + 
    \rho_{h, 0} \left( \frac{a}{a_{0}} \right)^{-3(1+\w_{h})}\right] \, \\
        & = \frac{\kappa}{3} \rho_0 \left( \frac{a}{a_{0}} \right)^{-3(1+\w)} 
        \left[1 + \frac{\rho_{h,0}}{\rho_{0}} \left(\frac{a}{a_{0}}\right)^{\frac{1 + \w_{h}}{1 + \w}}  \right] \, .
\end{align}
\end{subequations}
From relation \eqref{eq:Hlast}, we see that in addition to the dynamics generated by ordinary matter, we have also 
the energy density corresponding to hypermomentum which gives an additional contribution depending on the equation 
of state parameter $\w_{h}$, as expected. 
Quite remarkably, the form \eqref{rrt} for $\w_{h}=0$ coincides with the 
usual functional dependence that one assumes for the spin density in Weyssenhoff fluid models \cite{Medina:2018rnl}.

It is worth mentioning that in the case where neither of $\w$ and $\w_{h}$ is constant, we can still 
maintain the conventional form of the continuity equation if $\rho_{h}$ is related to $\rho$ through 
\beq\label{neg4}
\int \frac{d \rho_{h}}{(1+ \w_{h})\rho_{h}} = \int \frac{d \rho}{(1+ \w)\rho} + C \,.
\eeq

To give an example, if the hypermomentum component behaves as a generalized Chaplygin gas, i.e. 
$p_{h}=-\frac{A}{\rho_{h}^{n}}$,  ($ n \neq -1$) and the barotropic index $\w$ is constant we find the relation
\begin{equation}
 \rho_{h}(\rho)= \left[A+(\rho_{h,0}^{2}-A)\left(\frac{\rho}{\rho_{0}}\right)^{\frac{1+n}{1+\w}} 
 \right]^{\frac{1}{1+n}} \,.   
\end{equation}

On the other hand, if the hypermomentum part in \eqref{neg4} obeys a polytropic equation of state of the form 
$p_{h}=\alpha \rho_{h}^{1+\frac{1}{n}}$\,, $\alpha>0$, taking for instance, $n=2$ and for $\w=const.$ we get
for energy density
\beq
\rho_{h}(\rho)=\left(\frac{C\rho^{\frac{1}{2(1+\w)}}}{1-\alpha C\rho^{\frac{1}{2(1+w)}}}\right)^{2} \,.
\eeq
From the last formula, we see that for high energy densities, $\rho_{h}\sim const.$ and hypermomentum acts, effectively, \
as a cosmological constant. Of course, the above examples are only some possibilities and 
more fundamental relations for the equation of state parameter can only be obtained if the microscopic kinetic 
theory of the fluid is known.

\section{Conclusion and Outlook}
\label{sec: 7 Conclusions}

In metric-affine gravity both the metric and affine connection are taken to be independent variables, which introduces torsion and nonmetricity as the post-Riemannian part of geometry. Correspondingly, the matter sources in this setting are characterized by the energy-momentum as well as hypermomentum, arising from the variation of the matter action with respect to the metric and connection, respectively. In the context of cosmology, it is useful to model the matter as a cosmological hyperfluid, i.e.\ as a perfect fluid in terms of energy-momentum and taking the hypermomentum components to obey the same set of cosmological symmetries. Imposing the cosmological principle restricts the post-Riemannian part of the connection to five independent components, two in torsion and three in nonmetricity, while on the matter side there are two spin, one dilation, and two shear degrees of freedom in hypermomentum, as was worked out only recently \cite{Iosifidis:2020gth}. The connection equations relate these two sets to each other algebraically, thus in the absence of the hypermomentum sources the post-Riemannian part of the connection also vanishes, reducing the theory to general relativity.

The goal of the present paper was to deliver the first systematic investigation of cosmic hyperfluids, asking how would the presence of hypermomentum quantities like spin and shear affect the evolution of the universe. We studied the most straightforward metric-affine gravity model, where the usual Einstein--Hilbert action is computed from a generic connection. The projective invariance of that simple gravitational action makes the dilaton and the related nonmetricity component to vanish. In addition, one of the spin parts along with its respective torsion component trivializes to mimic spatial curvature, and we dropped them as well. Therefore, the system of the FLRW cosmological equations describes the evolution of the Hubble function sourced by the energy density and pressure, plus one spin and two shear components of the hypermomentum. There is also a single continuity equation that combines all the energy-momentum and hypermomentum quantities. 

Solving the system is impossible without making assumptions about the hyperfluid equation of state, i.e.\ how the matter variables relate to each other, in analogy to the usual equation of state in general relativity which writes pressure as a function of density. The ans\"{a}tze we employed involved taking the spin and/or shear to be proportional to the square root of the energy density (motivated on dimensional grounds), considering a constant ratio $\w_h$ of the effective pressure and density of the hypermomentum terms, assuming the energy density and pressure to obey their usual continuity equation separately, or postulating de Sitter expansion. In a couple of examples with the second shear component assuming a ``slow roll'' regime of the energy density was also a useful approximation, backed up by the explicit phase diagrams.

To our surprise we discovered, that despite a rather complicated form, the equations yielded meaningful analytic solutions, often admitting an easy interpretation as modifications of the standard solutions in general relativity. It is a well known result in Friedmann cosmology, that a single quantity, the matter equation of state parameter $\w$ (the barotropic index), determines both how the universe expands and how the matter dilutes, i.e.\ the time evolution of the Hubble function and the energy density. In a number of cases with the hyperfluid, we obtained the same forms for the evolution laws as in general relativity, only the value of the effective index $\weff$ that determined expansion and the value of the index $\w_\rho$ that determined the dilution of matter started to deviate from $\w$. Thus for single component hypermomentum cases, and assuming the effective density of hypermomentum remains small compared to the energy density, we were able to expressly quantify the hypermomentum effects as correction terms to the general relativity behavior. The other ans\"{a}tze, like demanding constant $\w_h$, also allow analytic integration but restrict the available solutions more, and do not necessarily yield a general relativity limit where corrections can be added. Furthermore, equipped with these insights we were able to investigate several simple cases of multiple hyperfluid components too, still finding ways how to integrate the system completely or by invoking reasonable approximations.

The general setup explored in this paper opens an interesting avenue for cosmological model building. To run a realistic cosmology, the Universe must be populated by several different species of particles, modelled as fluids. It is not inconceivable that the particles responsible for the dark sector may actually carry significant hypermomentum, or perhaps even some of the known particles might have a weak coupling to the independent connection and are thus endowed with hypermomentum as well. Both cases would then feature as hyperfluids in the mix of cosmological fluids. Having a basic grasp how the spin and shear components of hypermomentum affect the cosmological expansion ($\weff$) and dilution of the hyperfluid density ($\w_\rho$), it should not be hard to construct models for a bouncing universe, inflation, early dark energy, dark matter amplification (hypermomentum contribution that mimics matter density in cosmology), variable or constant dark energy, effective spatial curvature, or some other problem that needs to be addressed. If the hyperfluid consists of massive particles it can be used to address the phenomena that involve inhomogeneities like structure formation and dark matter, while a hyperfluid made of massless particles can model homogeneously occurring phenomena like inflation or dark energy. For instance, in the massless case it seems possible to construct a hypermomentum equation of state that would contribute to the universe expansion as a cosmological constant and at the same time make the energy density drop slower than radiation but faster than dust matter. Such component would be cosmologically significant in the brief period between the radiation and matter dominated eras, and thus at least in terms of the background evolution behave as early dark energy, which has been proposed to solve the $H_0$ tension in cosmology \cite{Kamionkowski:2022pkx}. Given a model, its parameters can then be fitted with the data.

There are a number of directions for future investigations that can build upon the current work. For example, one may consider spatially curved metrics, or more complex assumptions about the hypermomentum equation of state, which could give a more interesting role to the spin $\zeta$ component. We may add extra terms into the gravitational action like the parity violating Hojman-Holst term or higher order invariants in curvature, torsion, and nonmetricity, or introduce nonminimal couplings with scalar and other fields. Then the projection invariance would be broken and the dilaton would acquire opportunities for nontrivial cosmological behavior as well. On the other hand, a more systematic knowledge is welcome about which fundamental matter actions can give rise to the cosmological spin and shear, and which hypermomentum equations of state would be reasonable from the fundamental physics point of view. This could lead to the formulation of a hypermomentum analogue of energy conditions, that postulate physically realistic constraints on the relation between energy and pressure in the case of normal fluids.
In this work, we investigated how hypermomentum degrees of freedom affect cosmological evolution at the background level. As a next step, the analysis must be extended to include the evolution of cosmological perturbations and their effects.

\section*{Acknowledgments}

In this work, IA, LJ, MS were supported by the Estonian Research Council via the Center of Excellence 
``Foundations of the Universe'' (TK202U4), and DI was supported by the Estonian Research Council grant SJD14.


\bibliographystyle{utphys}
\bibliography{ref}

\appendix

\section{Relations between physical and  hypermomentum variables}
\label{sec: appendix}

In this appendix, we systematically present the relations between the hypermomentum degrees of freedom 
and the quantities for which we can give a physical meaning. 
The five degrees of freedom of the hypermomentum $\zeta$, $\phi$, $\chi$, $\psi$ and $\omega$ are described 
by the relation \eqref{hypermomentum:degrees}. 
It turns out that it is difficult to assign a physical meaning to these degrees of freedom directly. 
However, physical meaning can be given to quantities of dilation $\Delta$ \eqref{dilphiomega}, spin $\sigma$ \eqref{spinsig}, 
shear one $\Sigma_1$ and shear two $\Sigma_2$ \eqref{shearSig}. 
In the paper, we studied the dynamics of a homogeneous and isotropic universe, taking case-by-case some 
physical quantities ($\sigma$, $\Sigma_1$, $\Sigma_2$) to zero. Also, in our study the dilation $\Delta$ vanishes by default due to the projective invariance constraints on matter fields; see \eqref{condition:dilation:zero}. However, in this table, we keep the quantity $\Delta$ because it is non-vanishing when projective invariance is broken.
Note also that the $\zeta$-mode does not appear at all in our table. This is simply because this mode (being a pseudoscalar) does not interfere with the rest of the variables (see discussion around \eqref{eq zeta eq}).

The table \ref{variables:relations}  must, therefore, be read line by line, moving from left to right. The legend is as follows:
\begin{itemize}
\item{Vanishing variable: we assume that this physical variable is absent i.e.\ equal to zero (for $\Delta$, it is 
satisfied preserving the consistent approach, see \eqref{condition:dilation:zero}).}

\item{Additionally vanishing: we assume that the following physical variable is also vanishing, i.e.\ zero and moving 
from left to right,  we describe what this means for hypermomentum variables.}

\item{Condition(s): what the assumption made in the previous column implies for the hypermomentum variables.}

\item{Hypermomentum variables: what hypermomentum degrees of freedom are preserved when the constraint presented in the previous column is applied.}

\item{Relations between variables: what relations between hypermomentum degrees of freedom and physical quantities are preserved.}
\end{itemize}


\begin{table}[ht]
\centering
\begin{tabular} {|c|c|c|c|cc|}
\hline
    \multicolumn{1}{|c}{$\ $ Vanishing $\ $ } & \multicolumn{1}{|c}{\ Additionally \ } & \multicolumn{1}{|c}{$\quad$ Condition(s) $\quad$ } 
    & \multicolumn{1}{|c}{\ Hypermomentum \ } & \multicolumn{2}{|c|}{\ Relations between variables \ } \\
    \multicolumn{1}{|c}{variable} & \multicolumn{1}{|c}{vanishing} &\multicolumn{1}{|c}{} & \multicolumn{1}{|c}{variables} 
    &\multicolumn{2}{|c|}{} \\ 
    \hline 
    \hline 
     &  &  &  &  & \\
    $\Delta = 0$ & & $\omega = 3 \phi $ & $\phi, \chi, \psi$
    & $\sigma = \frac{1}{2} (\psi - \chi)$; \ $\Sigma_{1} = \frac{1}{2} (\psi + \chi)$; \ $\Sigma_{2} = \phi$ &   \\
    &  &  &  &  & \\
    & $\sigma = 0$  & $\psi = \chi$ & $\phi$, $\chi$ & $\Sigma_{1} = \chi$; \ $\Sigma_{2} = \phi$  & \\
    &  &  &  &  & \\
    & $\Sigma_{1} = 0$ & $\psi = - \chi$ & $\phi$, $\chi$ & $\sigma = - \chi$; \ $\Sigma_{2} = \phi$ & \\
    &  &  &  &  & \\
    & $ \Sigma_{2} = 0$ & $\phi = - \omega =0$ & $\chi$, $\psi$   & $\sigma = \frac{1}{2} (\psi - \chi)$; \ 
    $\Sigma_{1} = \frac{1}{2} (\psi + \chi)$ & \\
    &  &  &  &  & \\
    & $ \Sigma_{1} = \Sigma_{2} = 0$ & $\psi = - \chi,$ & $\chi$ &  $\sigma =  -\chi$ & \\
    & & $\phi = - \omega = 0$ & & & \\
    &  &  &  &  & \\
    \hline
    &  &  &  &  & \\
    $\sigma = 0$ &  & $\psi = \chi$ & $\phi$, $\chi$, $\omega$ & $\Delta = 3 \phi - \omega$; \ $\Sigma_{1} = \chi$; \ 
    $\Sigma_{2} = \frac{1}{4}(\phi + \omega)$ & \\
    &  &  &  &  & \\
    & $\Delta = 0$ & $\omega = 3 \phi$ & $\phi$, $\chi$ & $\Sigma_{1} = \chi$; \  $\Sigma_{2} = \phi$ & \\
    &  &  &  &  & \\
    & $\Sigma_{1} = 0$&  $\psi = - \chi = 0$ & $\phi$, $\omega$ & $\Delta = 3 \phi - \omega$; \, 
    $\Sigma_{2} = \frac{1}{4} (\phi + \omega)$ & \\
    &  &  &  &  & \\
    & $\Sigma_{2} = 0$&  $\phi = -\omega$ & $\phi$, $\chi$ & $\Delta = 4 \phi$; \ $\Sigma_{1} = \chi$ & \\
    &  &  &  &  & \\
    & $\Sigma_{1} = \Sigma_{2} = 0$&  $\psi = - \chi =0,$ & $\phi$ & $\Delta = 4 \phi$ & \\
    & & $\phi = -\omega$  & & & \\
    &  &  &  &  & \\
    \hline 
    &  &  &  &  & \\
    $\Sigma_{1} = 0$&  & $\psi = - \chi$ & $\phi$, $\chi$, $\omega$ & $\Delta = 3 \phi - \omega$; \ $\sigma = - \chi$; \
    $\Sigma_{2} = \frac{1}{4}(\phi + \omega)$  & \\
    &  &  &  &  & \\
    & $\Delta = 0$ & $\omega = 3 \phi$ & $\phi$, $\chi$   & $\sigma = -\chi$; \ $\Sigma_{2} = \phi$ & \\
    &  &  &  &  & \\
    &$\sigma = 0$  & $\psi = \chi = 0$ & $\phi$, $\omega$  & $\Delta = 3 \phi - \omega$; \ 
    $\Sigma_{2} = \frac{1}{4}(\phi + \omega)$ & \\
    &  &  &  &  & \\
    & $\Sigma_{2} =0$  & $\phi = - \omega$  & $\phi$, $\chi$  & $\Delta = 4 \phi$; \ $\sigma = - \chi$ & \\
    &  &  &  &  & \\
    \hline 
    &  &  &  &  & \\
    $\Sigma_{2} = 0$ &  & $\phi = - \omega$  & $\phi$, $\chi$, $\psi$  & 
    $\Delta= 4 \phi$; \ $\sigma= \frac{1}{2}(\psi - \chi)$; \ $\Sigma_{1} = \frac{1}{2}(\psi + \chi)$ & \\
    &  &  &  &  & \\
    & $\Delta = 0$ & $\omega = 3 \phi = 0$  & $\chi$, $\psi$  & $\sigma = \frac{1}{2} (\psi - \chi)$; \ $\Sigma_{1} = \frac{1}{2} (\psi + \chi)$ & \\
    &  &  &  &  & \\
    & $\sigma = 0$ & $\psi = \chi$  & $\chi$, $\psi$ & $\Delta = 4 \phi$; \ $\Sigma_{1} = \chi$ & \\
    &  &  &  &  & \\
    & $\Sigma_{1} = 0$ & $\psi = - \chi$  & $\phi$, $\chi$ & $\Delta = 4 \phi$; \ $\sigma = - \chi$ & \\
    &  &  &  &  & \\
    \hline 
    &  &  &  &  & \\
    $\Sigma_{1} = \Sigma_{2} = 0$& & $\psi = - \chi$,  & $\phi$, $\chi$   & $\Delta = 4 \phi$; \ $\sigma = -\chi$ & \\
    &  & $\phi = - \omega$  &  &  & \\
    &  &  &  &  & \\
    & $\Delta = 0$ & $\omega = 3 \phi = 0$ &  $\chi$ &  $\sigma = - \chi$ & \\
    &  &  &  &  & \\
    & $\sigma = 0$ & $\psi = \chi = 0$  & $\phi$ & $\Delta = 4 \phi$ & \\
    &  &  &  &  & \\
    \hline 
    \end{tabular}
    \caption{In this table, we summarize the relationships between the physical quantities $\Delta$, $\sigma$, 
    $\Sigma_1$, $\Sigma_2$ and the hypermomentum degrees of freedom  $\phi$, $\chi$, $\psi$, $\omega$. The legend of how to read 
    the table is given in the description below.}
    \label{variables:relations}
\end{table}

\end{document}